\documentclass[aps,prb,preprint,superscriptaddress,showpacs]{revtex4}
\usepackage{hyperref}
\usepackage{amsmath,amssymb,amsfonts}
\usepackage{graphicx}
\DeclareGraphicsExtensions{.jpg,.pdf,.png,.eps}
\newcommand{\BaP} {BaFe$_{2}$As$_{2}$ }
\newcommand{\BaPf} {BaFe$_{2}$As$_{2}$}
\newcommand{\BaRu} {Ba(Fe$_{1-x}$Ru$_{x}$)$_{2}$As$_{2}$ }
\newcommand{\BaRuf} {Ba(Fe$_{1-x}$Ru$_{x}$)$_{2}$As$_{2}$}
\newcommand{\Tc} {$T_{c}$ }
\newcommand{\Tcf} {$T_{c}$}
\newcommand{\Tsm} {$T_{sm}$ }
\newcommand{\Tsmf} {$T_{sm}$}
\newcommand{\BaRuNine} {Ba(Fe$_{0.91}$Ru$_{0.09}$)$_{2}$As$_{2}$ }
\newcommand{\BaRuNinef} {Ba(Fe$_{0.91}$Ru$_{0.09}$)$_{2}$As$_{2}$}
\newcommand{\BaRuOneSix} {Ba(Fe$_{0.84}$Ru$_{0.16}$)$_{2}$As$_{2}$ }
\newcommand{\BaRuOneSixf} {Ba(Fe$_{0.84}$Ru$_{0.16}$)$_{2}$As$_{2}$}
\newcommand{\BaRuTwoOne} {Ba(Fe$_{0.79}$Ru$_{0.21}$)$_{2}$As$_{2}$ }
\newcommand{\BaRuTwoOnef} {Ba(Fe$_{0.79}$Ru$_{0.21}$)$_{2}$As$_{2}$}
\newcommand{\BaRuTwoEight} {Ba(Fe$_{0.72}$Ru$_{0.28}$)$_{2}$As$_{2}$ }
\newcommand{\BaRuTwoEightf} {Ba(Fe$_{0.72}$Ru$_{0.28}$)$_{2}$As$_{2}$}

\begin{document}
\title{Combined effects of pressure and Ru substitution on BaFe$_{2}$As$_{2}$}
\author{S. K. Kim}
\affiliation{Ames Laboratory, Iowa State University, Ames, Iowa 50011, USA}
\affiliation{Department of Physics and Astronomy, Iowa State University, Ames, Iowa 50011, USA}
\author{M. S. Torikachvili}
\affiliation{Department of Physics and Astronomy, Iowa State University, Ames, Iowa 50011, USA}
\affiliation{Department of Physics, San Diego State University, San Diego, California 92182, USA}
\author{E. Colombier}
\affiliation{Ames Laboratory, Iowa State University, Ames, Iowa 50011, USA}
\author{A. Thaler}
\affiliation{Ames Laboratory, Iowa State University, Ames, Iowa 50011, USA}
\affiliation{Department of Physics and Astronomy, Iowa State University, Ames, Iowa 50011, USA}
\author{S. L. Bud'ko}
\affiliation{Ames Laboratory, Iowa State University, Ames, Iowa 50011, USA}
\affiliation{Department of Physics and Astronomy, Iowa State University, Ames, Iowa 50011, USA}
\author{P. C. Canfield}
\affiliation{Ames Laboratory, Iowa State University, Ames, Iowa 50011, USA}
\affiliation{Department of Physics and Astronomy, Iowa State University, Ames, Iowa 50011, USA}
\date{\today}

\begin{abstract}
The $ab$-plane resistivity of Ba(Fe$_{1-x}$Ru$_{x}$)$_{2}$As$_{2}$ ($x$~=~0.00, 0.09, 0.16, 0.21, and 0.28) was studied under nearly hydrostatic pressures, up to 7.4~GPa, in order to explore the $T-P$ phase diagram and to compare the combined effects of iso-electronic Ru substitution and pressure.  The parent compound BaFe$_{2}$As$_{2}$ exhibits a structural/magnetic phase transition near 134~K.  At ambient pressure, progressively increasing Ru concentration suppresses this phase transition to lower temperatures at the approximate rate of $\sim$~5~K/\%Ru and is correlated with the emergence of superconductivity.  By applying pressure to this system, a similar behavior is seen for each concentration: the structural/magnetic phase transition is further suppressed and superconductivity induced and ultimately, for larger $x$~Ru and $P$, suppressed.  A detailed comparison of the $T-P$ phase diagrams for all Ru concentrations shows that 3~GPa of pressure is roughly equivalent to 10\% Ru substitution.  Furthermore, due to the sensitivity of Ba(Fe$_{1-x}$Ru$_{x}$)$_{2}$As$_{2}$ to pressure conditions, the melting of the liquid media, 4~:~6 light mineral oil~:~n-pentane and 1~:~1 iso-pentane~:~n-pentane, used in this study could be readily seen in the resistivity measurements.  This feature was used to determine the freezing curves for these media and infer their room temperature, hydrostatic limits: 3.5 and 6.5~GPa, respectively.

\end{abstract}

\maketitle 
\section{Introduction}
Many studies have investigated the effects of electron, hole, and isovalent substitutions in the AEFe$_{2}$As$_{2}$ (AE=Alkaline Earth) system.\cite{Canfield10,Rotter08,Rotter10,Kasahara10,Kim10,Ni08,Ni10,Sefat08,Sefat09,Fisher09,Thaler10}  For \BaPf , in some cases this substitution causes the suppression of the structural/magnetic transition temperature (\Tsmf) and the emergence of superconductivity.\cite{Canfield10,Ni08,Ni10,Rotter08,Rotter10,Kasahara10,Kim10}  In other cases, such as substitution of Cr or Mn for Fe, \Tsm is suppressed without superconductivity ever stabilizing.\cite{Kim10,Sefat09}  For the case of \BaRuf , increasing the concentration of isovalent Ru,\cite{Thaler10} reveals behavior similar to Co substitutions,\cite{Ni08} but without introducing additional charge carriers into the system.\cite{Thaler10,Rullier10}  Pressure has also been used as an iso-electronic tuning mechanism.\cite{Colombier09,Matsubayashi09,Ishikawa09,Duncan10,Yamazaki10,Torikachvili08,Torikachvili08a,Alireza09,Kotegawa09}  As pressure increases, \Tsm is suppressed gradually and disappears at the critical pressure, $P_{crit}$.  Near $P_{crit}$ the superconducting temperature (\Tcf) reaches its maximum value and the transition width is narrowest.\cite{Colombier09,Colombier10}  Although the ambient pressure $T-x$ phase diagram for \BaRu and $T-P$ phase diagram for \BaP manifest similar features and can be scaled to each other,\cite{Thaler10} at ambient pressure and with increasing Ru concentration, \BaRu undergoes an increase in the lattice parameter a and a decrease in c which causes an overall increase in volume.  The potential similarities between the effects of hydrostatic pressure and Ru substitution are intriguing.  In order to better quantify and understand the similarities between the effects of Ru substitution and pressure, in this study, we determine the $T-P$ phase diagrams for multiple Ru substitution levels and explore the possibility of a universal scaling between these isoelectronic tuning mechanisms.  

In addition, it is well known that the behavior of \BaP is sensitive to pressure conditions.\cite{Colombier09,Matsubayashi09,Ishikawa09,Duncan10}  Pressure inhomogeneities associated with non-hydrostatic conditions tend to decrease the pressure needed to suppress \Tsm and induce superconductivity.  This sensitivity causes discrepancies in the construction of the $T-P$ phase diagram depending on the pressure conditions.  Therefore, conditions as close to hydrostatic as possible are necessary for consistent results.  In this study, a piston-cylinder cell and a modified Bridgman cell with appropriate liquid media were used to measure the resistivity of \BaRu samples under pressure.  A maximum pressure of 7.4~GPa was achieved.  Although parent \BaP has already been measured several times under various pressure conditions,\cite{Colombier09,Matsubayashi09,Ishikawa09,Duncan10,Yamazaki10} it was measured again under the same conditions as the rest of the \BaRu samples in order to allow more reliable comparisons as well as to gauge the level of hydrostaticity of the liquid medium.  We find a remarkably simple scaling between pressure and Ru substitution: 3~GPa of applied pressure affects the phase diagram in a manner similar to 10$\%$ Ru substitution for Fe. 

\section{Experimental methods}

All \BaRu single crystals measured in this study were grown out of self flux using the method described elsewhere.\cite{Thaler10}  

Electrical resistivity measurements under pressures of up to 2.3~GPa were conducted using a piston cylinder pressure cell.\cite{Torikachvili08,Torikachvili08a}  Higher pressures, up to 7.4~GPa, were achieved using a Bridgman cell that has been modified to work with liquid pressure media.\cite{Colombier07}  Both these cells were designed to fit inside a Quantum Design Physical Properties Measurement System (PPMS) which served as a variable temperature station for the temperature range between 1.8 and 300~K.  

The piston-cylinder cell has a Be-Cu body with a center core made out of tungsten carbide. The samples for this cell were cut into rectangles with dimensions of approximately 1.5~$\times$~0.3~$\times$~0.1~mm$^{3}$.  Four Pt wires were attached to the sample using Epotek H20E silver loaded epoxy.  The feedthrough containing the sample, Manganin, and Pb manometers was inserted in a polytetrafluoroethylene cup containing a 4~:~6 mixture of light mineral oil and n-pentane, which served as the liquid pressure-transmitting medium, unless otherwise stated.  Pressure was applied at ambient temperature with a hydraulic press, using the Manganin as a reference manometer.  A calibrated Cernox sensor was attached to the body of the cell for temperature measurements.  At low temperatures, the pressure was determined from the superconducting temperature, \Tcf , of the Pb manometer.\cite{Eiling81}  The cooling and warming rates were kept below 0.5~K/min which corresponded to a temperature lag between the sample and Cernox sensor of less than 0.5~K at high temperatures and less than 0.1~K for temperatures less than 70~K.  Further details are already described elsewhere.\cite{Torikachvili08a}  Cooling data are shown in this work unless otherwise stated.  

The modified Bridgman cell has a Be-Cu body with opposing, non-magnetic, tungsten-carbide anvils.  A 1~:~1 mixture of iso-pentane~:~n-pentane was used as the liquid pressure-transmitting medium.  Although we determine that this liquid medium has a higher hydrostatic limit of 6.5~GPa (see Appendix)(hydrostatic limit being defined as the pressure at which the medium begins to solidify at ambient temperature) than Fluorinert mixtures,\cite{Sidorov05,Piermarini73,Sakai99,Klotz09} it also has a higher compressibility which means lower maximum pressures can be achieved without changes to critical cell dimensions.  Moreover, there was a higher rate of failure for the wires within the sample chamber when using the iso-pentane~:~n-pentane mixture, where a wire would break or otherwise lose contact with the sample.  Despite these difficulties, the higher hydrostatic limit made it preferable over other liquid media (e.g. Fluorinert mixtures with hydrostatic limits in the 1-2~GPa range).  For the Bridgman cell, samples were cleaved and cut into approximately 700~$\times$~150~$\times$~30~$\mu$m$^{3}$ and four 12.5~$\mu$m diameter gold wires were spot welded onto the sample to create electrical contacts for standard four-probe measurements.  The pressure within the cell was determined using the superconducting temperature, \Tcf , of Pb.\cite{Eiling81}  For these cells, the difference between the pressure at room temperature and at low temperature was previously determined to be less than 0.1~GPa.\cite{Colombier10}  For all Bridgman cell measurements, data that were taken while warming from base temperatures.  For T~$<$~35~K data were taken after the temperature was stabilized at each point, ensuring a minimal thermal gradient between the cell and the sample.  For measurements above 35~K, the cell was warmed at a rate of 0.5~K/min which leads to a maximum temperature lag of approximately 1.2~K.\cite{Colombier10}  Only warming data are shown in this work. 

Due to the small dimensions of the samples used in the Bridgman cell, resistivity values can have errors of up to 50\%.  Furthermore, the micaceous nature of the crystals makes them prone to exfoliation, a tendency which is compounded by the inevitable damage inflicted by the cleaving and cutting done to shape them into the appropriate dimensions.  Great care was taken to choose samples with the fewest of these defects, but it is possible that under pressure, the layers could be compressed or further distorted, leading to changes in the strains in the sample resulting in small jumps or changes in resistivity values.  So as to provide a better view of the evolution of the sample behavior with pressure, the piston cylinder cell data were normalized so that the ambient temperature and pressure resistivity value matched that of the corresponding Bridgman cell sample.   

\begin{figure}[!ht]
\begin{center}
\includegraphics[angle=0,width=100mm]{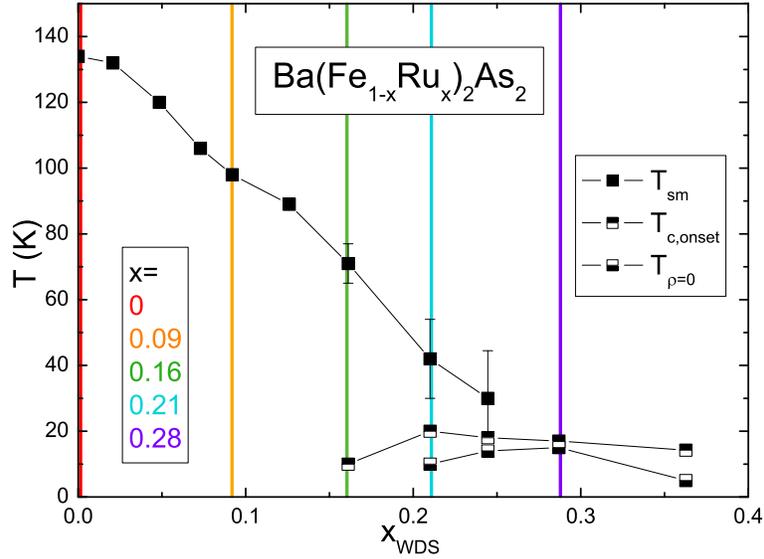}
\end{center}
\caption{(Color online) $T-x$ phase diagram for \BaRuf .\cite{Thaler10}  The vertical lines indicate the Ru concentrations that were studied under pressure.}
\label{TpPD}
\end{figure}

Figure \ref{TpPD} shows the $T-x$ phase diagram for \BaRuf .\cite{Thaler10}  The vertical lines indicate the Ru concentrations that were chosen for this study in order to explore the low-$x$ and optimal-$x$ regions of the phase diagram.  The onset and offset criteria for \Tc are shown in Fig.~\ref{Ba122}(b).  The onset \Tc was taken as the intersection of the extrapolated lines seen in the inset of Fig.~\ref{Ba122}(b).  The offset \Tc was taken as the temperature at which the resistivity reaches zero as seen in Fig.~\ref{Ba122}(b), denoted as $T_{c,\rho =0}$. 

Strain induced, granular/filamentary superconductivity is known to occur in many of the AEFe$_{2}$As$_{2}$ systems.\cite{Torikachvili08a,Kotegawa09,Saha09,Hu11}  To gauge the impact of this effect on the superconducting phase transition, current dependent resistivity measurements were done at various pressures.  Figure \ref{currentdep} shows two such measurements.  At 3.64~GPa (Fig.~\ref{currentdep}(a)) only the onset of the superconducting transition is seen and there is a definite dependence on the applied current which suggests that granular/filamentary superconductivity is responsible for the resistance decrease.  At 6.21~GPa (Fig.~\ref{currentdep}(b)) this current dependent behavior is less prominent, but still seen during the superconducting transition.  The difference in the offset temperature of the superconducting transition between 0.01~mA and 1~mA of applied current is $\sim$~3~K.  So as to minimize the effects of granular/filamentary superconductivity, a 1~mA current was used for all measurements.  

\section{Results}
\subsection{\texorpdfstring{BaFe$_{2}$As$_{2}$}{space}}
Previous pressure measurements of \BaP with a modified Bridgman cell have been reported \cite{Colombier09,Ishikawa09} using a Fluorinert (FC) mixture of 1~:~1 FC70 and FC77 as the liquid pressure-transmitting medium.  The hydrostatic limit for this medium is $\sim$~1~GPa,\cite{Sakai99} thus an additional, poorly-controlled, small, uniaxial stress component is likely at higher pressures.  Due to the sensitivity of \BaP to uniaxial stress, a different liquid medium, a 1~:~1 mixture of iso-pentane~:~n-pentane, with a higher hydrostatic limit of 6.5~GPa (see Appendix) was used in this study.  

\begin{figure}[!ht]
\begin{center}
\includegraphics[angle=0,width=100mm]{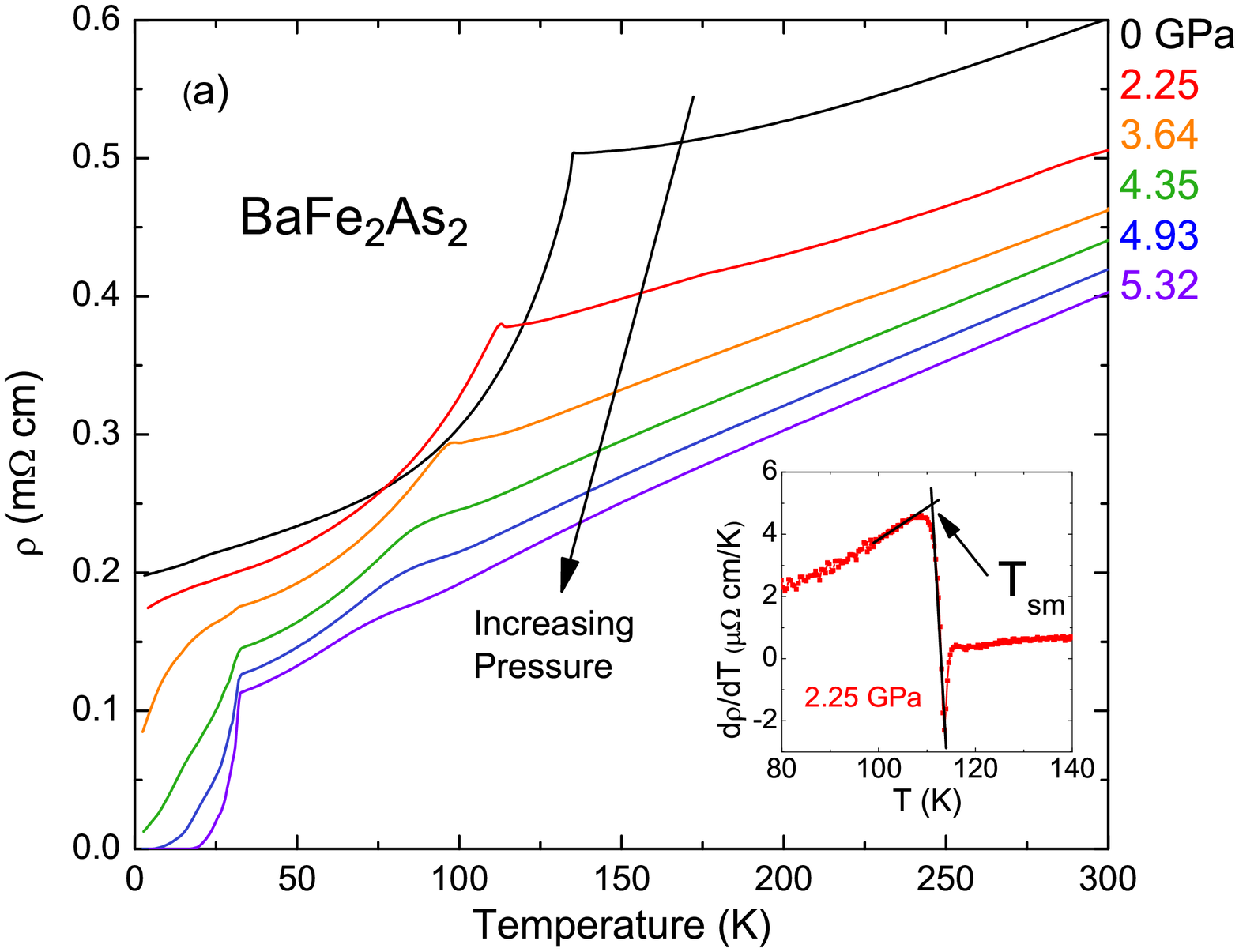}
\includegraphics[angle=0,width=100mm]{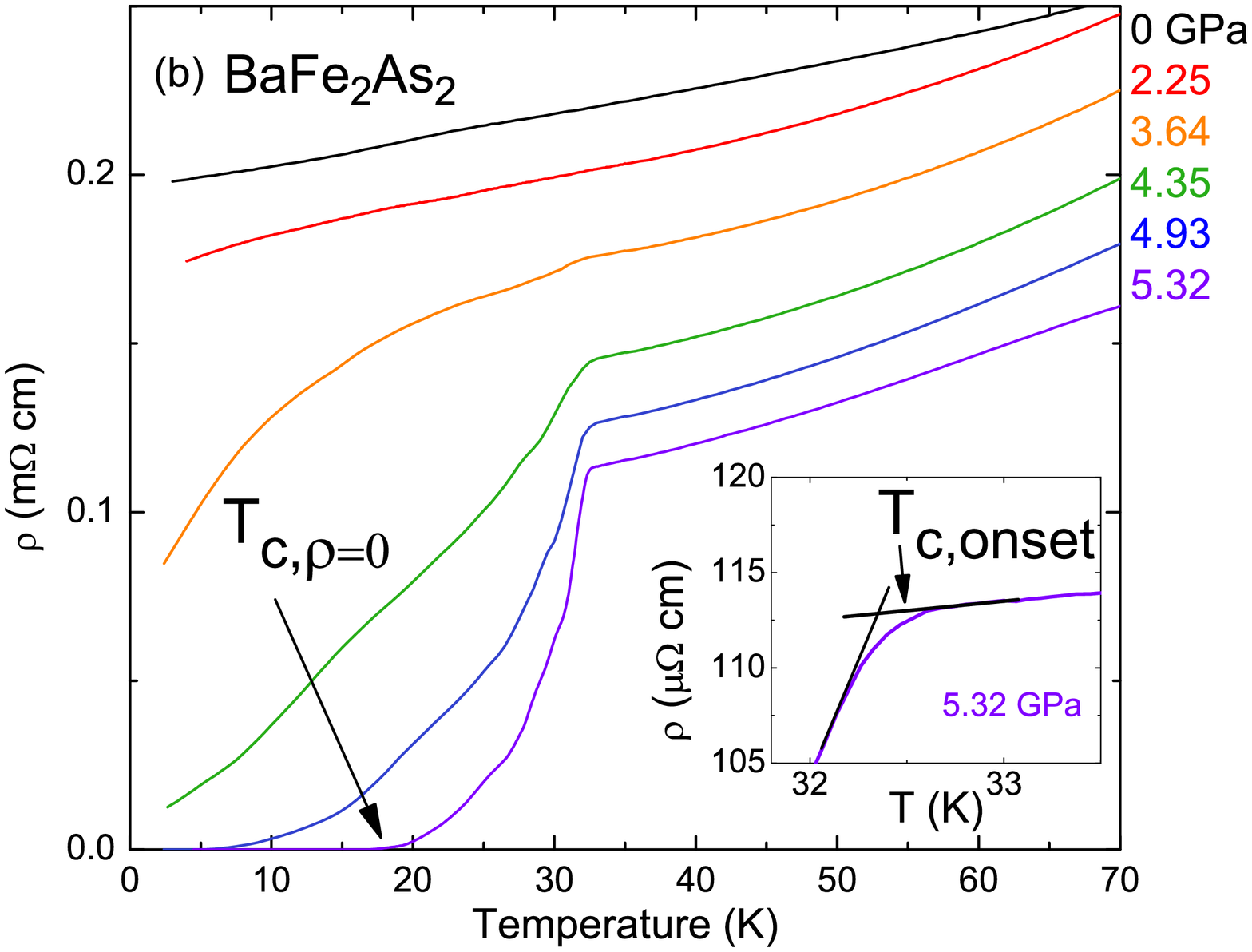}
\end{center}
\caption{(Color online) Temperature dependence of the resistivity of \BaP for pressures up to 5.32~GPa measured using the modified Bridgman cell.  (a)~Measurements are shown for temperatures up to 300~K.  Inset shows criteria used for the determination of \Tsmf . (b)~Same measurements shown for temperatures up to 70~K with criterion used for $T_{c,\rho =0}$.  Inset shows the criterion used for $T_{c,onset}$.}
\label{Ba122}
\end{figure}

\begin{figure}[!ht]
\begin{center}
\includegraphics[angle=0,width=100mm]{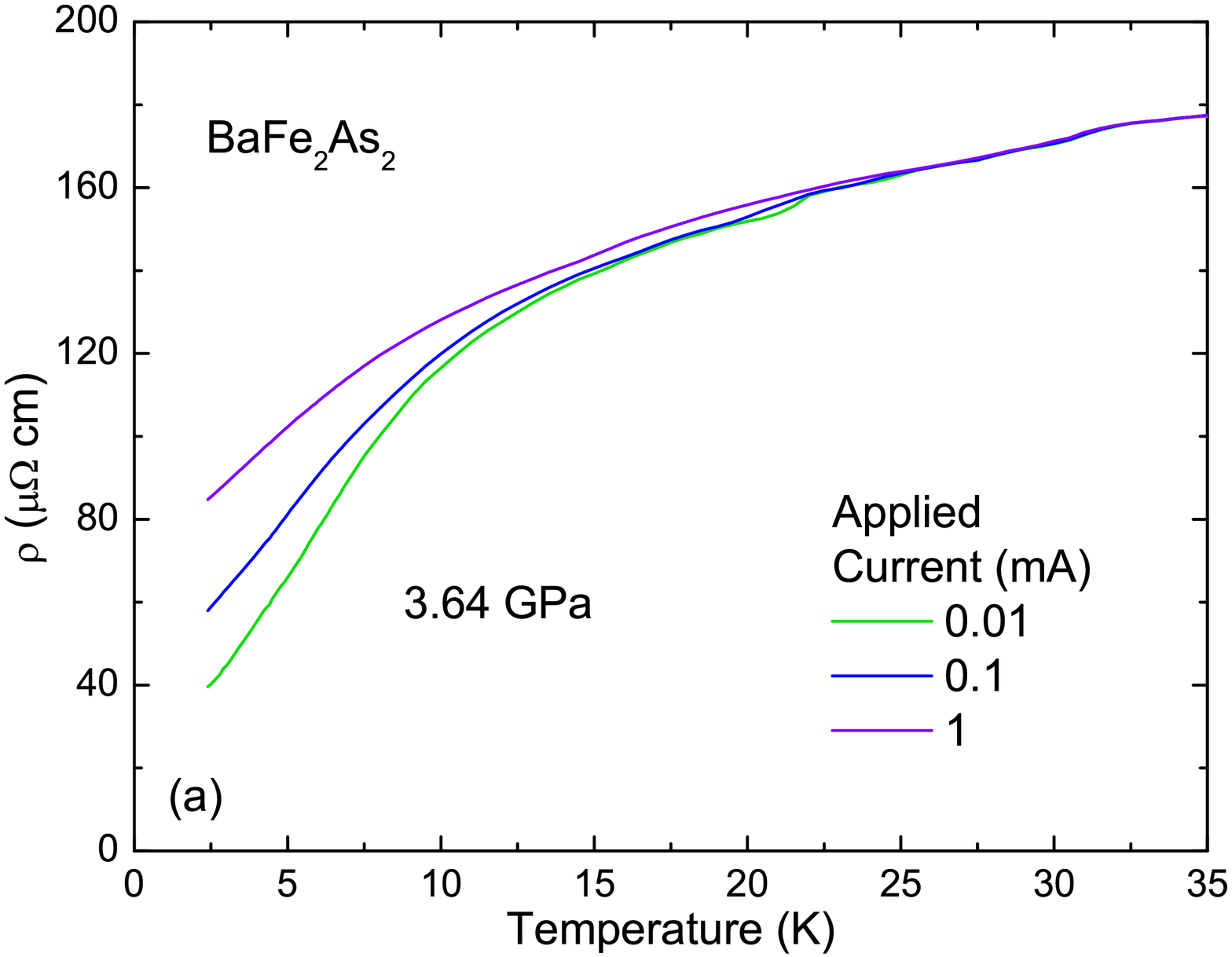}
\includegraphics[angle=0,width=100mm]{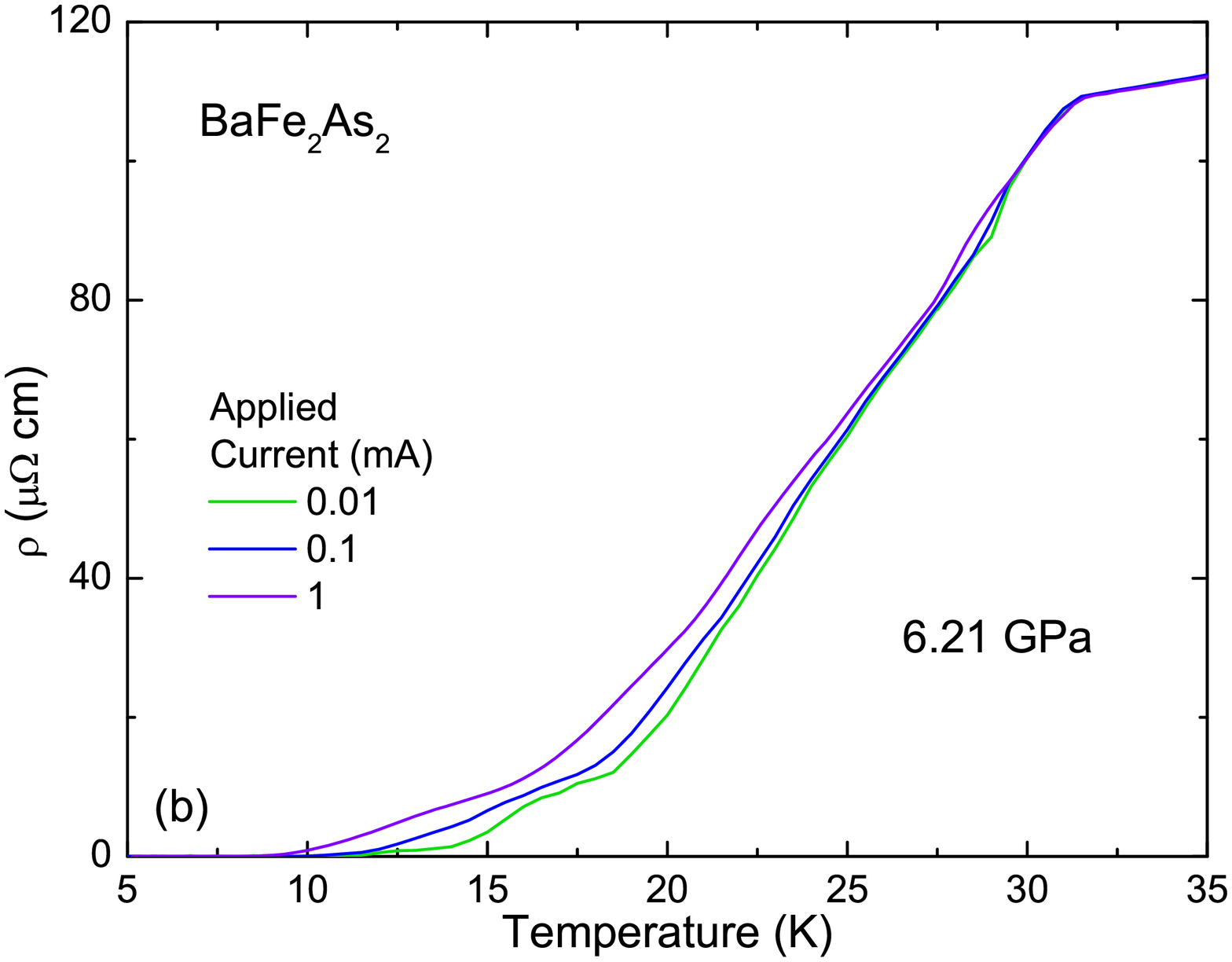}
\end{center}
\caption{(Color online) Resistivity measurements of \BaP  with applied currents of 0.01, 0.1, and 1 mA at (a)~3.64~GPa and (b)~6.21~GPa.}
\label{currentdep}
\end{figure}

Two samples of \BaP were measured using the Bridgman cell; one measured up to 5.32~GPa (Fig.~\ref{Ba122}) and the other up to 6.71~GPa (not shown).  The ambient pressure resistivity of \BaP decreases on cooling.  At $\sim$~134~K, the sample undergoes a structural/magnetic transition where it converts from a high temperature tetragonal, paramagnet to a low temperature, orthorhombic, antiferromagnet.  As pressure is applied, the resistivity decreases and the structural/magnetic transition moves to lower temperatures and broadens.  In addition, a small downturn arises at low temperature and as a precursor to the superconducting transition.  This increasingly kink-like feature is reminiscent of the pressure induced, granular/filamentary, superconducting behavior of SrFe$_{2}$As$_{2}$\cite{Colombier09} and CaFe$_{2}$As$_{2}$.\cite{Torikachvili08a}  A current dependent resistivity measurement at 3.64~GPa (Fig.~\ref{currentdep}(a)) suggests that superconductivity in a small fraction of the sample, most likely due to internal strains, precedes the occurrence of a more robust superconducting state, when $\rho(T)$ is much less sensitive to the excitation current, as shown in Fig.~\ref{currentdep}(b).

\begin{figure}[!ht]
\begin{center}
\includegraphics[angle=0,width=100mm]{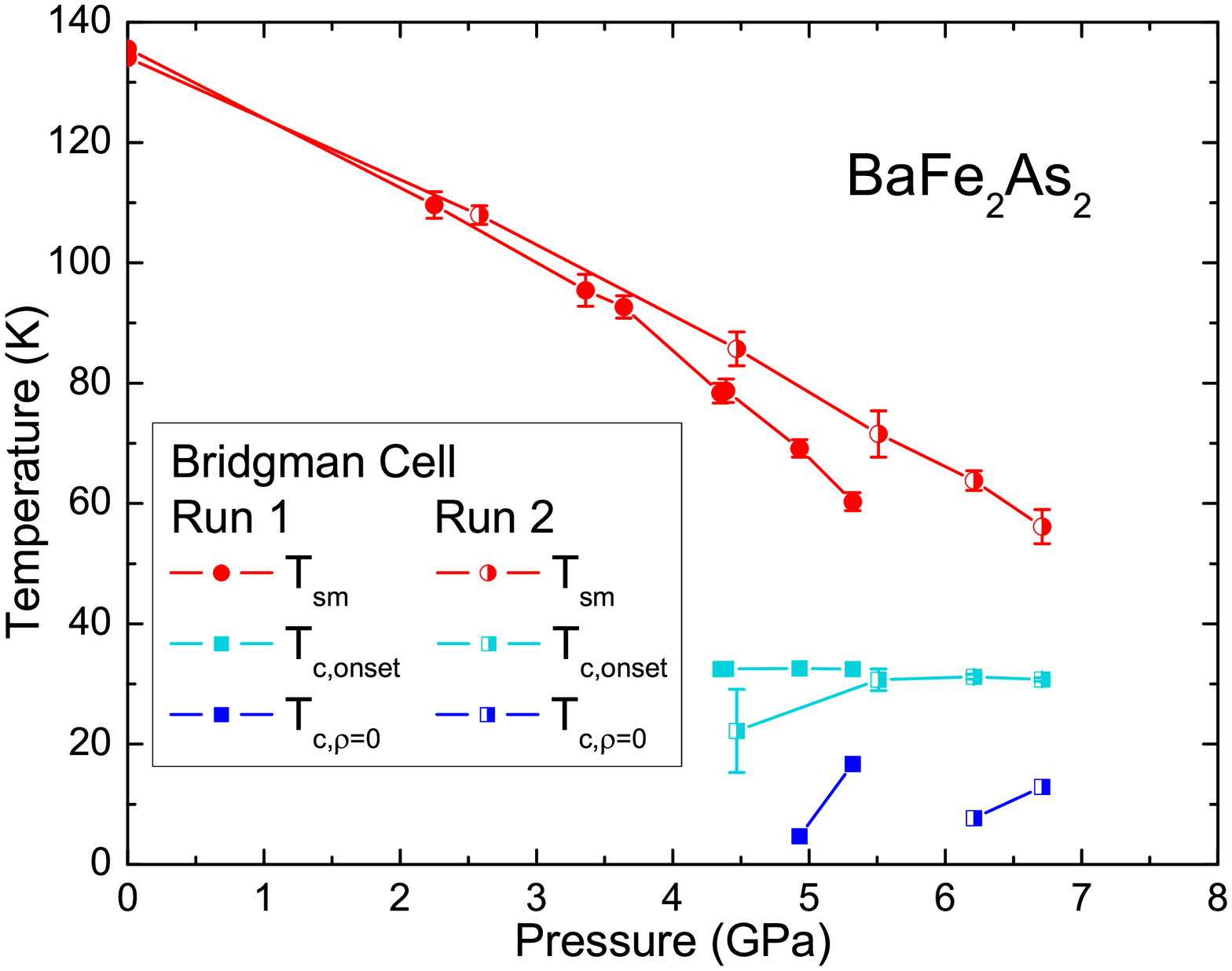}
\end{center}
\caption{(Color online) $T-P$ Phase Diagram for two sets of measurements of \BaP using the Bridgman cell.  The iso-pentane~:~n-pentane mixture was used as the liquid, pressure transmitting medium for both measurements.}
\label{Ba122PD}
\end{figure}

The resistive feature associated with the structural/magnetic transition is gradually suppressed with pressure but still persists at the maximum pressure achieved (6.4~GPa) even with the emergence of a finite $T_{c,\rho =0}$ at $\sim$~5~GPa.  The structural/magnetic transition temperatures for \BaP were taken as the maximum of the derivative of the resistivity as seen in the inset of Fig.~\ref{Ba122}(a).  The general form of the phase diagram is not very dependent on the hydrostaticity of the pressure, however, the features in the phase diagram shift towards higher pressure as hydrostaticity is improved.\cite{Colombier09,Matsubayashi09,Ishikawa09,Duncan10,Yamazaki10}  The resulting phase diagram for pure \BaP using the iso-pentane~:~n-pentane mixture is shown in Fig.~\ref{Ba122PD}.  The phase diagrams for the two separate measurements show qualitatively similar behavior with a quantitative shift of about 1.5 GPa in the transition temperatures at the highest pressures for Run 2.  Unfortunately, it is these last three, highest pressure, data points that are associated with manometer inconsistencies that may be associated with over estimating the actual pressure experienced by the sample.  The phase diagram presented in Fig. \ref{Ba122PD} is in qualitative agreement with previous measurements of \BaP under pressure in a Bridgman cell using the Fluorinert mixture,\cite{Colombier10} but with all transition temperatures shifted to higher pressures for the iso-pentane~:~n-pentane mixture.  

\subsection{\texorpdfstring{Ba(Fe$_{0.91}$Ru$_{0.09}$)$_{2}$As$_{2}$}{space}}
As shown in Fig.~\ref{Ba122PD}, for pure \BaPf , superconductivity is just being stabilized in the $P~\sim$~5~GPa range, while the resistive signature of the structural/magnetic transition remains visible up to our highest measured pressures.  For the first Ru concentration in this study, we chose $x$~=~0.09 which has no bulk superconductivity and an approximately 35~K suppression of \Tsm ($\sim$~98~K) from that of parent \BaP (\Tsm =~134 K).

\begin{figure}[!ht]
\begin{center}
\includegraphics[angle=0,width=80mm]{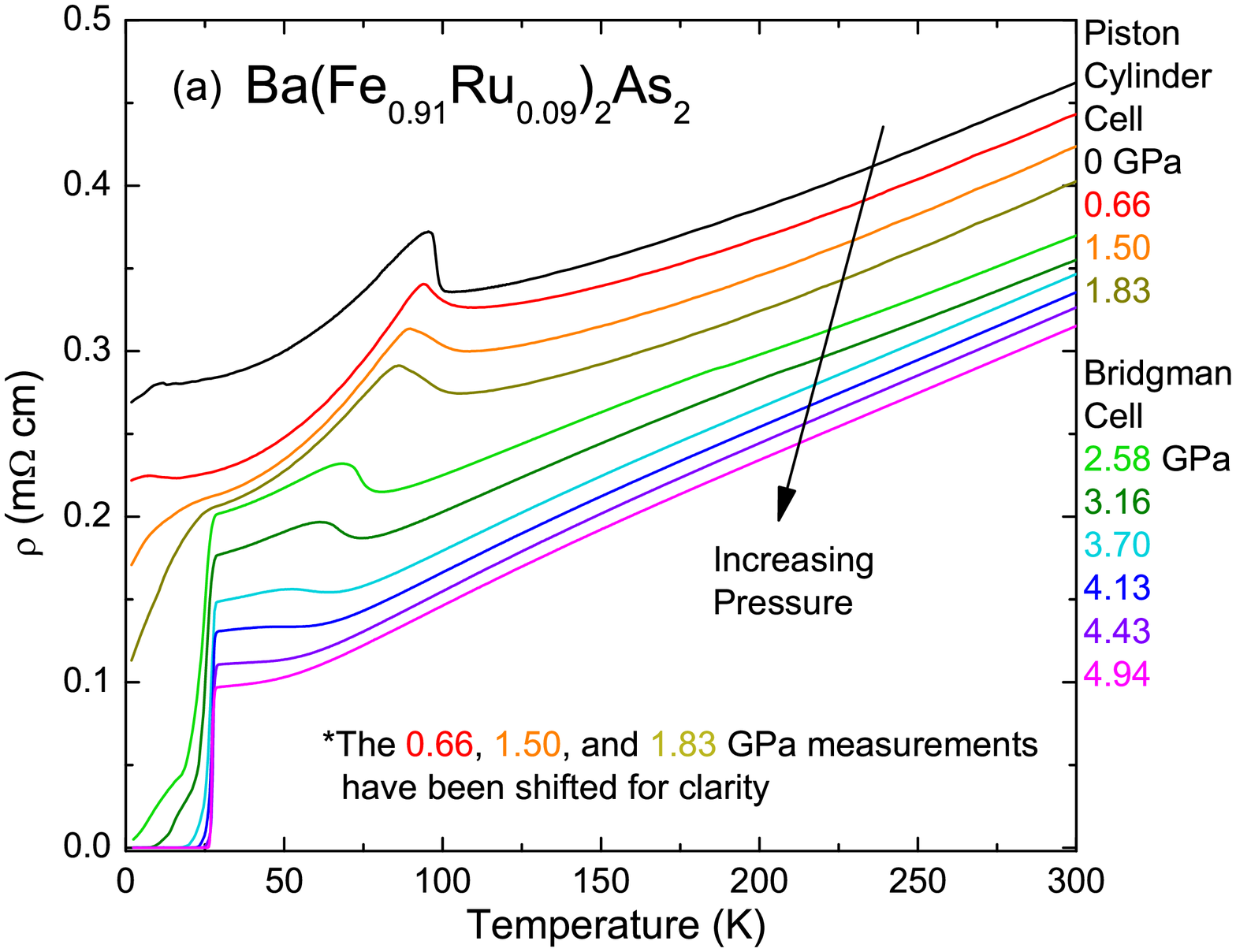}\\
\includegraphics[angle=0,width=80mm]{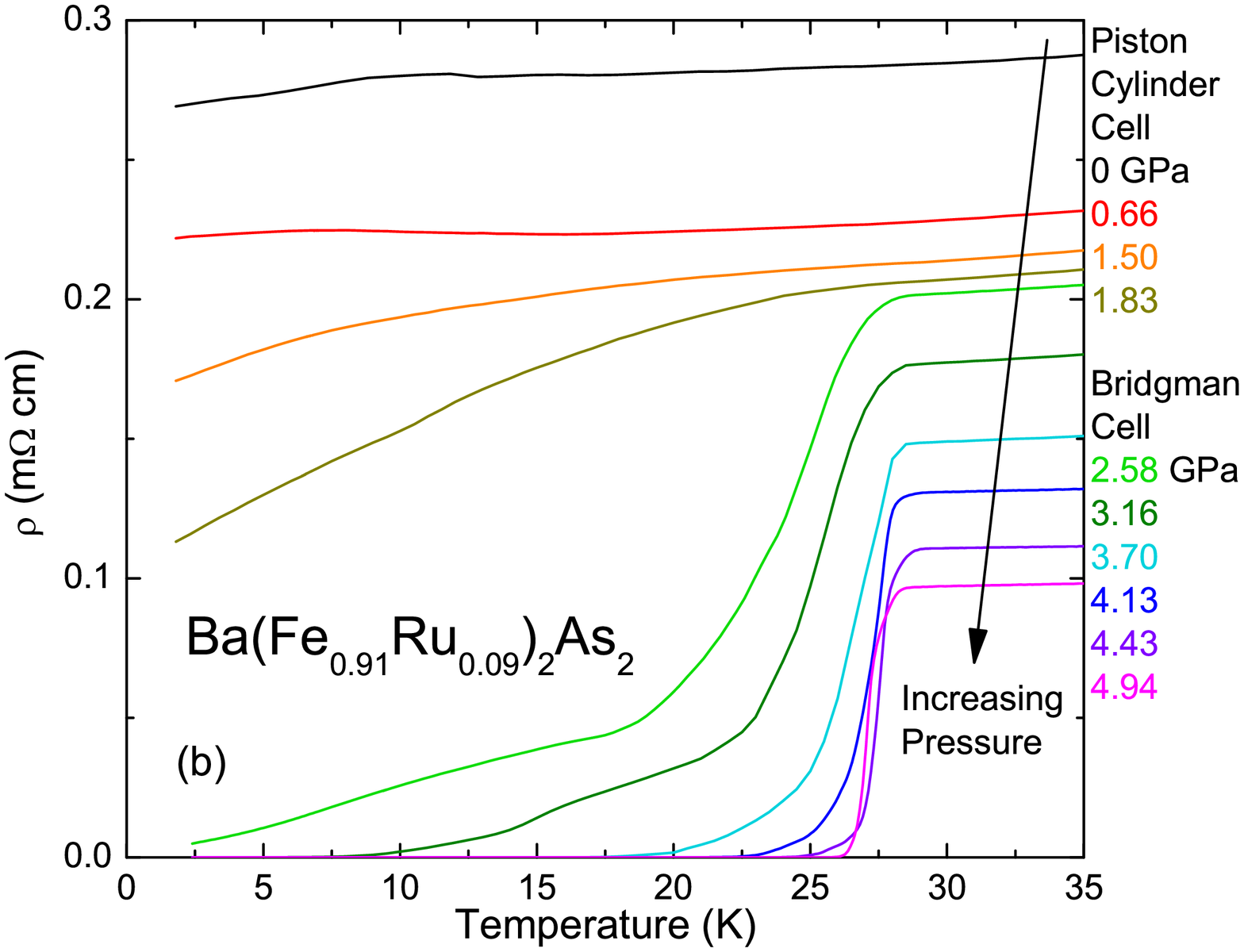}\\
\includegraphics[angle=0,width=40mm]{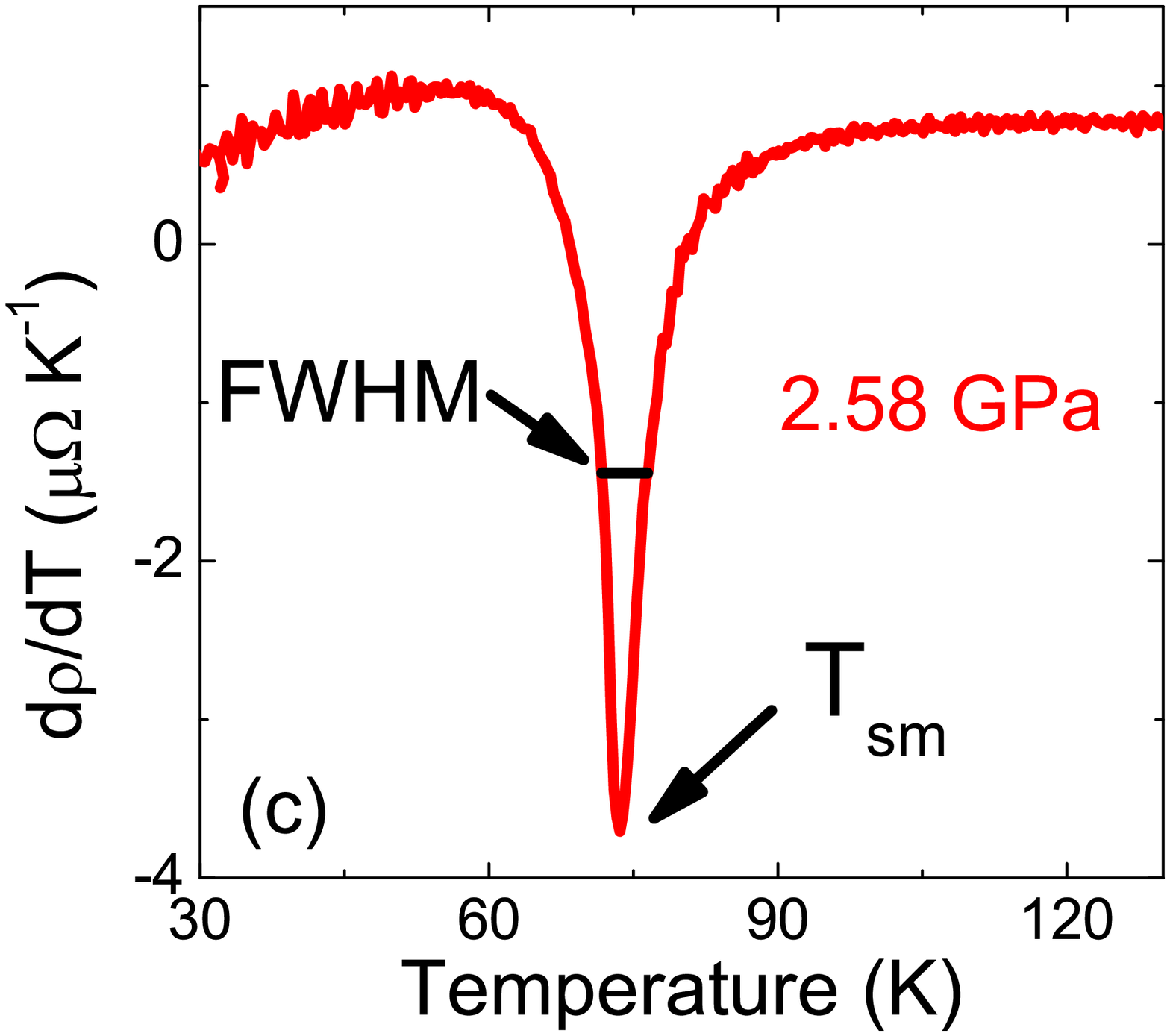}
\includegraphics[angle=0,width=40mm]{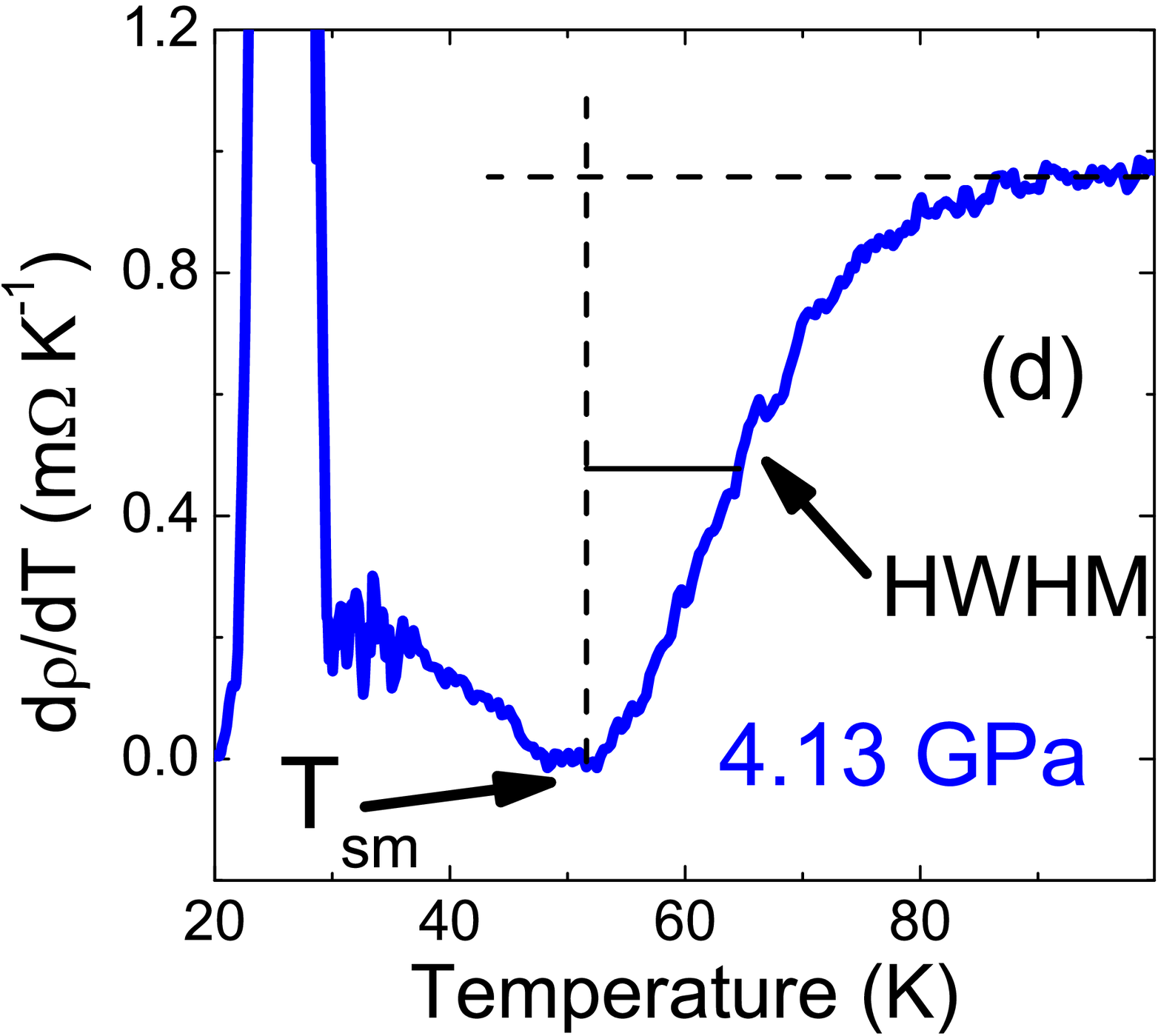}
\end{center}
\caption{(Color Online) Temperature dependence of the resistivity of \BaRuNine up to 1.83~GPa and 4.94~GPa using a piston cylinder cell and a Bridgman cell, respectively.  (a)~Shown for temperatures up to 300~K.  The 0.66, 1.50, and 1.83~GPa measurements have been shifted down by 0.035, 0.045, and 0.06~m$\Omega$ cm, respectively, for clarity. (b)~Shown for temperatures up to 35~K. (c)~and~(d)~Criteria used to determine \Tsm and their corresponding error bars.}
\label{Ru9}
\end{figure}

Two samples of \BaRuNine were measured: one with the piston cylinder cell up to 1.83~GPa and another with the Bridgman cell up to 4.94~GPa (Fig.~\ref{Ru9}). Figure \ref{Ru9} shows the effects of pressure on the resistivity of \BaRuNine samples.  

With increasing pressure, \Tsm is gradually suppressed to lower temperatures and granular/filamentary superconductivity develops and gradually shorts out more of the sample.  When zero resistivity is achieved, with 3.16~GPa of pressure, a small feature due to the structural/magnetic transition can still be observed, suggesting that the suppression of the structural/magnetic transition is not complete.  Further pressure increase almost completely suppressed the structural/magnetic transition and increased $T_{c,\rho =0}$ to a value of 25.7~K at $P_{crit}=$~4.94~GPa.  The superconducting transition width also decreased with pressure.  At this critical pressure, there is no measurable current dependence of the resistivity curve, suggesting the development of bulk superconductivity.  As will be seen for higher Ru substitutions, these features are consistent with $P_{crit}\approx$~5~GPa for this sample.

\begin{figure}[!ht]
\begin{center}
\includegraphics[angle=0,width=90mm]{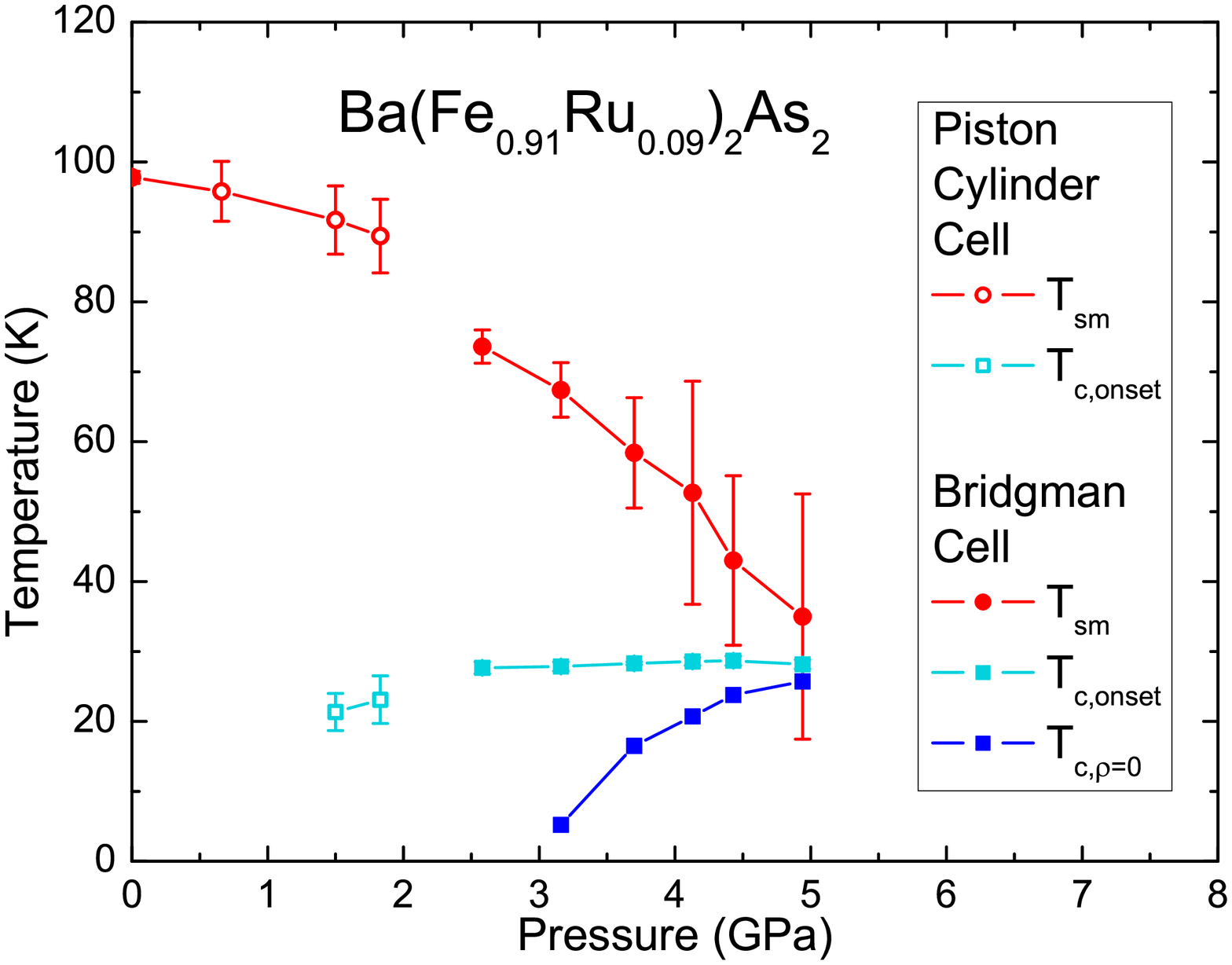}
\end{center}
\caption{(Color Online) $T-P$ Phase Diagram for measurements of \BaRuNinef .  Open and solid symbols indicate measurements using the piston cylinder cell and the modified Bridgman cell, respectively.}
\label{Ru9PD}
\end{figure}

A phase diagram constructed from these measurements can be seen in Fig.~\ref{Ru9PD}.  For all Ru substituted samples that were measured, \Tsm was taken as the minimum of the resistivity derivative (with error bars coming from the width at half maximum), which can be seen in Fig.~\ref{Ru9}(c) and \ref{Ru9}(d).  As \Tsm is suppressed, the minimum of $d\rho/dT$, which was used to determine \Tsm becomes broader and near $P_{crit}$, becomes indistinguishable from the onset of \Tcf .  The phase diagram shows a consistent qualitative behavior with \Tsm decreasing with pressure and a superconducting \Tc dome arising at higher pressures; the addition of $x$~=~0.09 Ru simply shifts $P_{crit}$ and the superconducting dome to lower pressures.  

\subsection{\texorpdfstring{Ba(Fe$_{0.84}$Ru$_{0.16}$)$_{2}$As$_{2}$}{space}}
\begin{figure}[!ht]
\begin{center}
\includegraphics[angle=0,width=90mm]{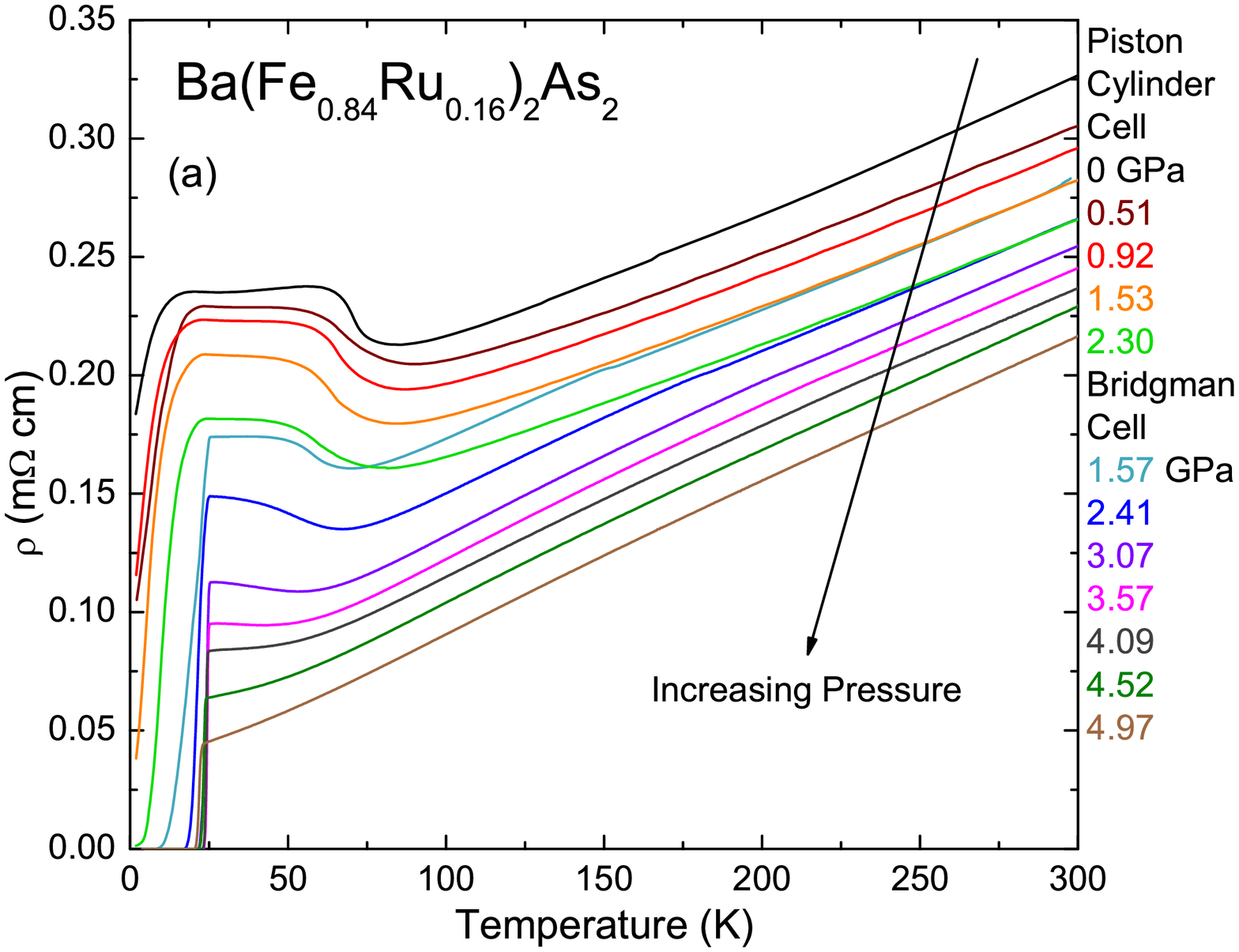}
\includegraphics[angle=0,width=90mm]{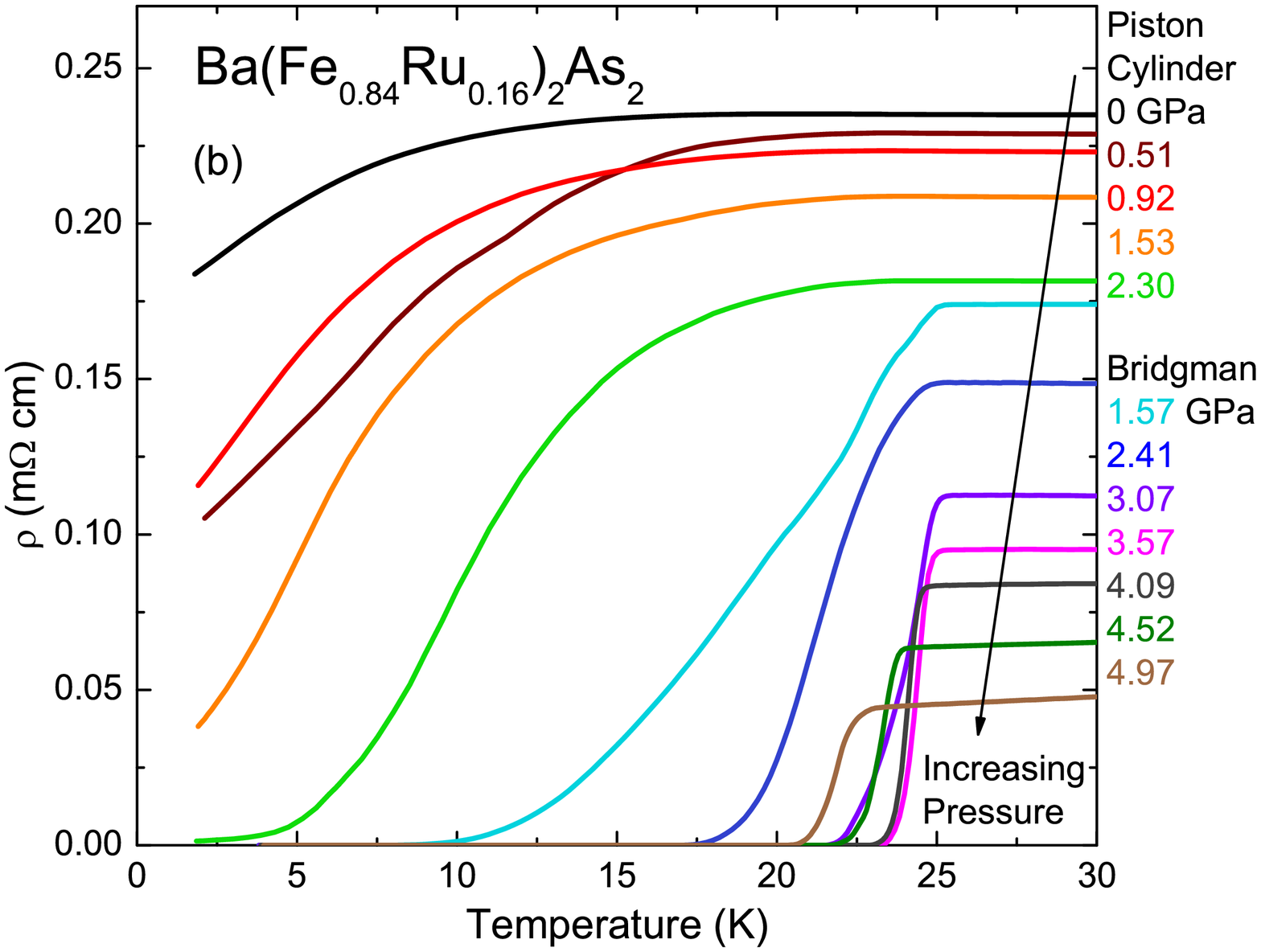}
\end{center}
\caption{(Color online) Temperature dependence of the resistivity of \BaRuOneSix up to 2.30 and 4.97~GPa measured using the piston cylinder cell and the modified Bridgman cell, respectively.  (a)~Shown for temperatures up to 300~K. (b)~Shown for temperatures up to 30~K.}
\label{Ru16}
\end{figure}

Measurements of $\rho(T,P)$ were carried out on three samples of \BaRuOneSixf : one with the piston cylinder cell up to 2.30~GPa and two with the Bridgman cell with maximum pressures of 1.57~GPa and 4.97~GPa. As shown in Fig.~\ref{TpPD}, \BaRuOneSix also resides on the low-$x$ side of the $T-x$ phase diagram, but with a further reduction of the structural/magnetic phase transition and much closer proximity to the superconducting dome.  Ambient pressure resistivity measurement (Fig.~\ref{Ru16}) of \BaRuOneSix  shows both the structural/magnetic transition as well as the onset of superconductivity.  Added pressure decreases \Tsm and a finite $T_{c,\rho =0}$ is achieved with 1.57~GPa of pressure.  A maximum $T_{c,\rho =0}$ of 23~K is achieved with 3.57~GPa and the narrowest superconducting transition width is realized at 4.09~GPa with a $T_{c,\rho =0}$ of 22.9~K and width of $\Delta T_{c}$~$\sim$~0.4~K.  At 4.09~GPa, the structural/magnetic transition has all but disappeared.  Further pressure increase causes the structural/magnetic transition to disappear completely, a decrease in \Tcf , and a broadening of the superconducting transition.  

\begin{figure}[!ht]
\begin{center}
\includegraphics[angle=0,width=90mm]{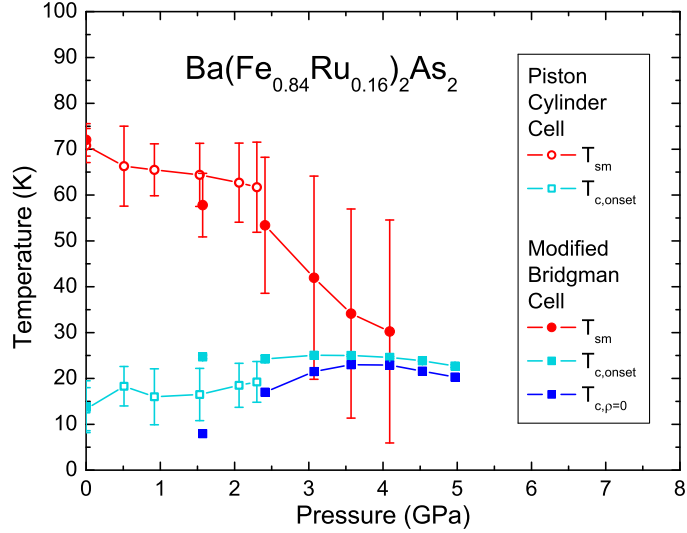}
\end{center}
\caption{(Color online) $T-P$ Phase Diagram for measurements of \BaRuOneSixf .  Open and solid symbols indicate measurements using the piston cylinder cell and the modified Bridgman cell, respectively.}
\label{Ru16PD}
\end{figure}

For the low pressure, piston cylinder cell measurements, the structural/magnetic transition at 0.5~GPa is broader than at 0.9~GPa and $T_{c,onset}$ is also higher.  One possible cause of this is that the first pressurization could have caused strains in the sample due to a small increase in pressure from constrictions and contractions of the cell from the first cooling and warming of the cell.  Of greater concern is the fact that there are noticeable differences between measurements done in the piston cylinder and the Bridgman cell.  For the 1.53~GPa and 2.30~GPa measurements from the piston-cylinder cell and 1.57 and 2.41~GPa measurements from the Bridgman cell, the corresponding sets of the temperature dependent resistivity data overlap well from room temperature down to $\sim$~150~K, below which the resistivity of the sample in the Bridgman cell is suppressed much faster.  Furthermore, in this pressure range, the Bridgman cell measurements manifest a sharp superconducting transition whereas, for the piston cylinder cell, the transition is wider and does not reach $\rho$~=~0 even at the base temperature of 1.8~K.  In addition, the $T_{c,onset}$ values are consistently lower in the piston cylinder cell than in the Bridgman cell and the rate of suppression of \Tsm is smaller in the piston cylinder cell.  These differences suggest a slight disparity in the degree of hydrostaticity between the Bridgman cell, using the iso-pentane~:~n-pentane, and the piston cylinder cell with the light mineral oil~:~n-pentane mixture.  In the $x$~=~0.09~Ru measurements, these differences were also seen, although smaller.  Despite these discrepancies, the combined phase diagram shown in Fig.~\ref{Ru16PD} demonstrates rather good agreement between measurements taken with these two cells.

\subsection{\texorpdfstring{Ba(Fe$_{0.79}$Ru$_{0.21}$)$_{2}$As$_{2}$}{space}}
\begin{figure}[!ht]
\begin{center}
\includegraphics[angle=0,width=100mm]{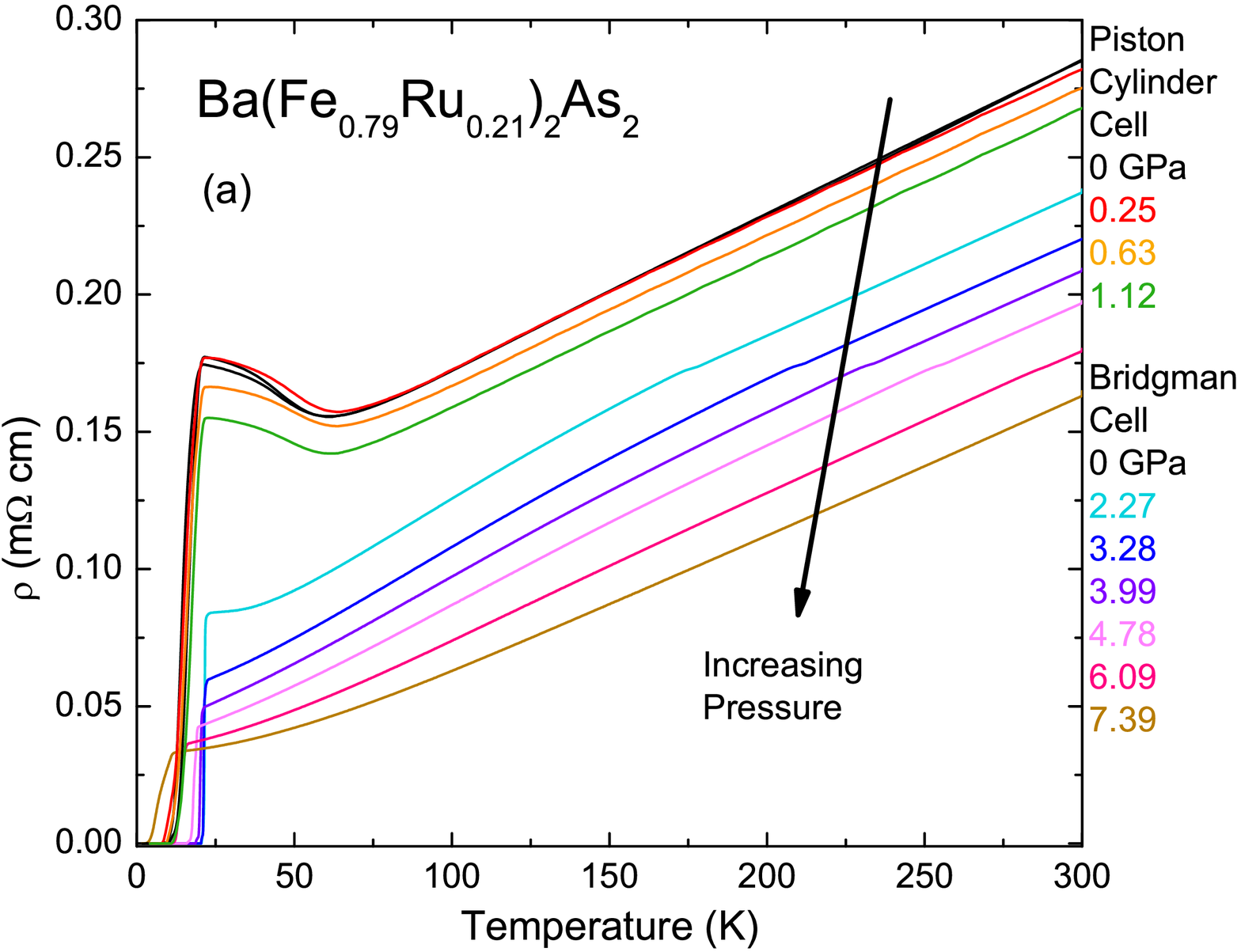}
\includegraphics[angle=0,width=100mm]{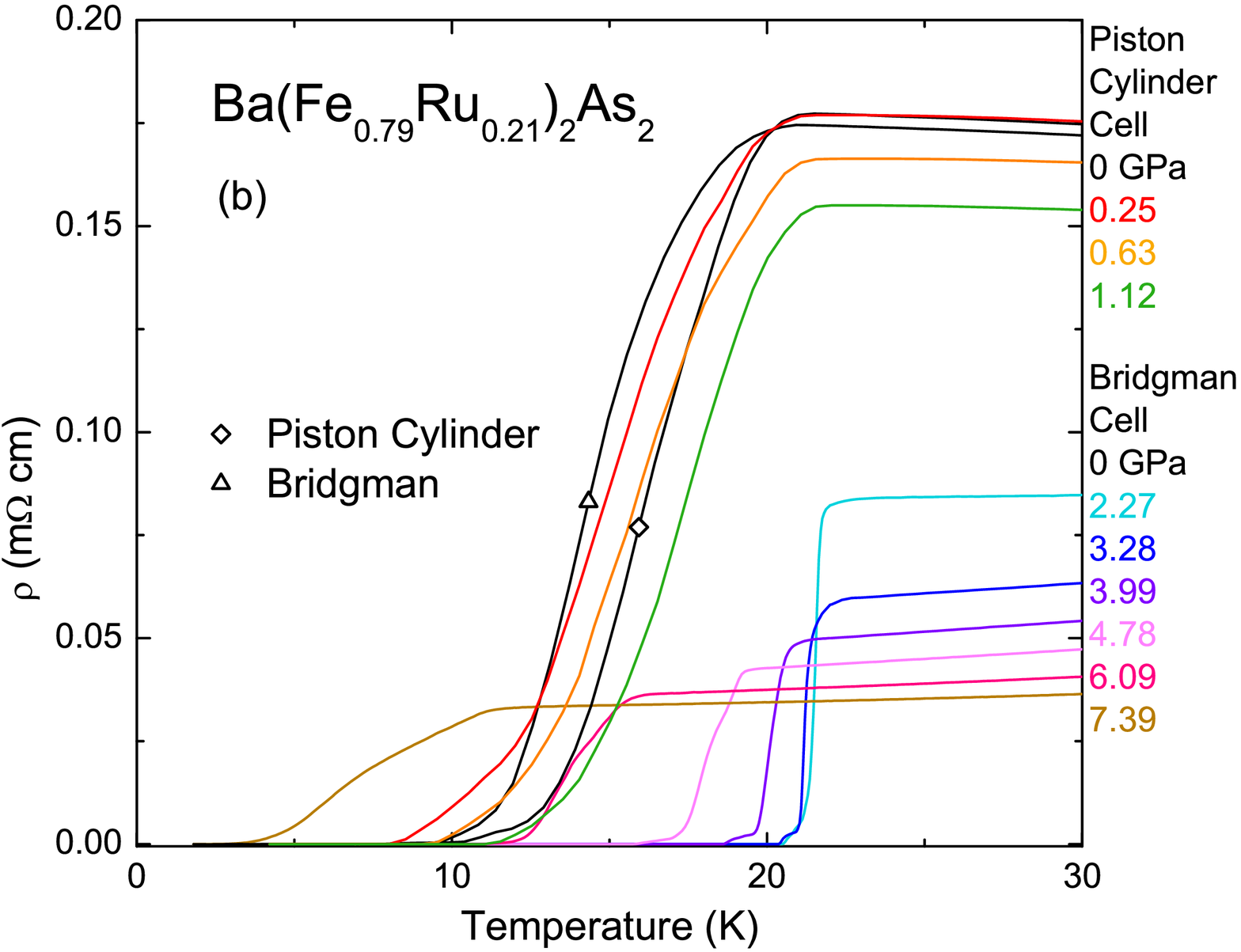}
\end{center}
\caption{(Color online) Temperature dependence of the resistivity of \BaRuTwoOne up to 1.12 and 7.39~GPa measured using the piston cylinder cell and the modified Bridgman cell, respectively. (a)~Shown for temperatures up to 300~K. (b)~Shown for temperatures up to 30~K.}
\label{Ru21}
\end{figure}

\BaRuTwoOne is very close to the optimal Ru concentration (see Fig.~\ref{TpPD}).  At the higher Ru concentrations, the homogeneity of the Ru substitution starts to vary within the batch of samples, as reported by Thaler, \textit{et al.}\cite{Thaler10}  Figure~\ref{Ru21} shows the results of resistivity measurements for the samples used in the piston cylinder cell and the Bridgman cell for pressures up to 1.12 and 7.39~GPa, respectively, both using the iso-pentane~:~n-pentane mixture.  At ambient pressure, \BaRuTwoOne samples show a coexistence of both the structural/magnetic transition and superconductivity.  The ambient pressure \Tc for the two samples used in the cells differ by $\sim$~1~K.  A maximum $T _{c,\rho =0}$ of 20.3~K was achieved with only 2.27~GPa and also has the narrowest transition width at this pressure.  Further pressure increases causes the suppression of \Tc and a widening of the transition width.  

\begin{figure}[!ht]
\begin{center}
\includegraphics[angle=0,width=100mm]{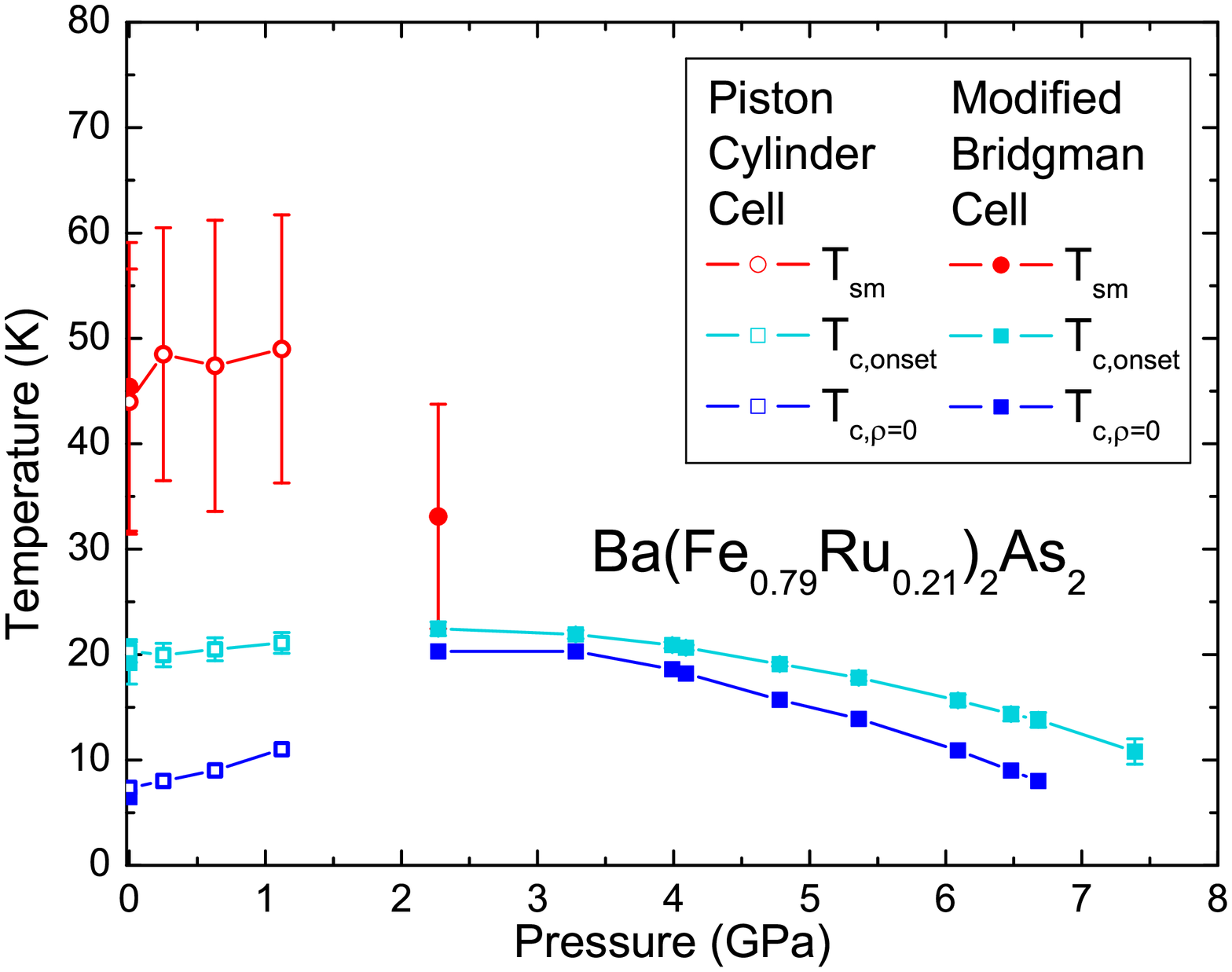}
\end{center}
\caption{(Color online) $T-P$ Phase Diagram for measurements of \BaRuTwoOnef .  Open and solid symbols indicate measurements using the piston cylinder cell and the modified Bridgman cell, respectively.}
\label{Ru21PD}
\end{figure}

The phase diagram for \BaRuTwoOne is shown in Fig.~\ref{Ru21PD}.  \Tsm in the piston cylinder cell ($P$~$<$~1.2~GPa) is only weakly affected by pressure, whereas by $P$~=~2.27~GPa (the first finite pressure in the Bridgman cell) \Tsm was significantly decreased.  As with other substitution levels, \Tcf$(P)$ forms a dome-like region with the highest and sharpest \Tc found near $P_{crit}$~=~3.28~GPa.  

\subsection{\texorpdfstring{Ba(Fe$_{0.72}$Ru$_{0.28}$)$_{2}$As$_{2}$}{space}}
\begin{figure}[!ht]
\begin{center}
\includegraphics[angle=0,width=80mm]{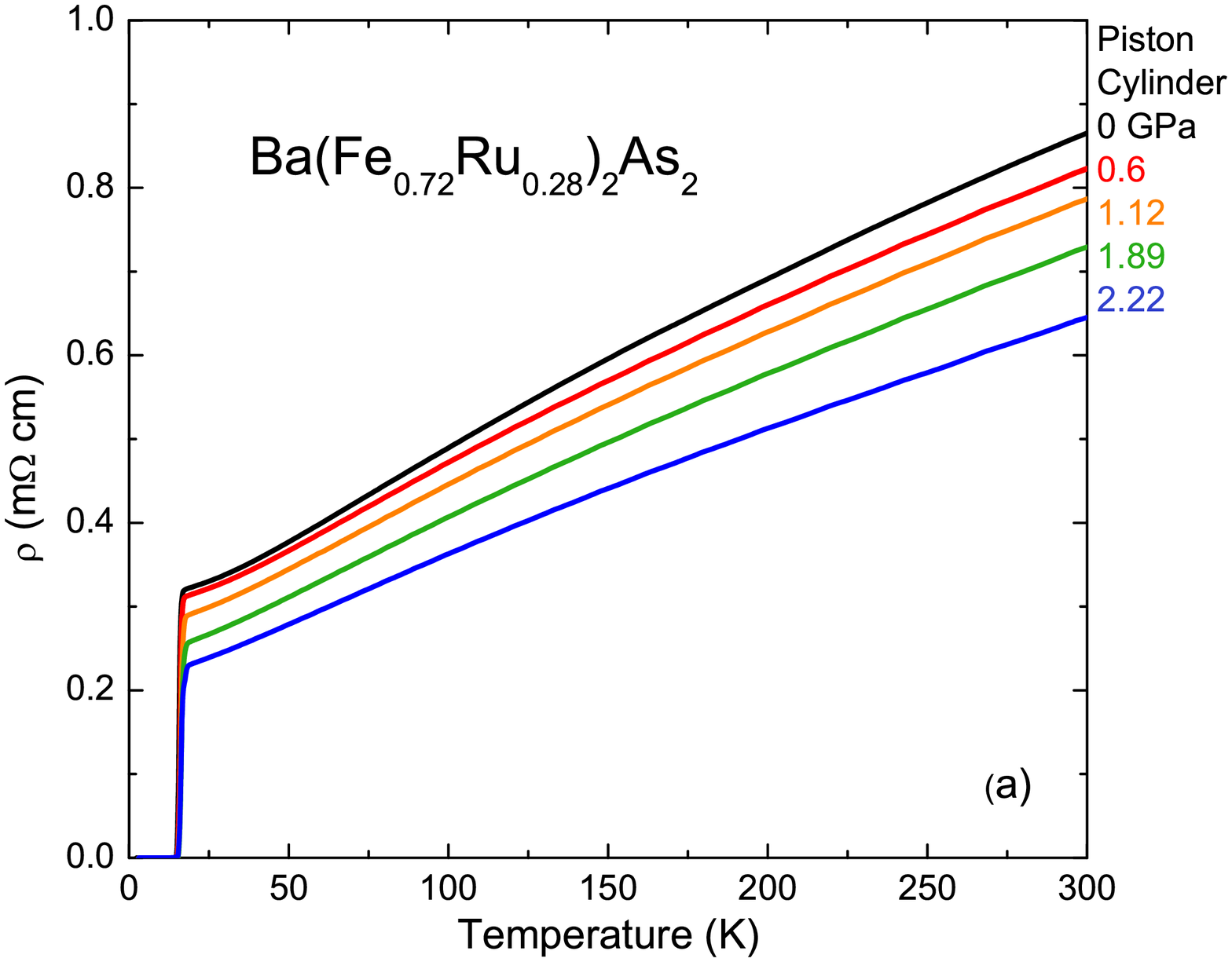}
\includegraphics[angle=0,width=80mm]{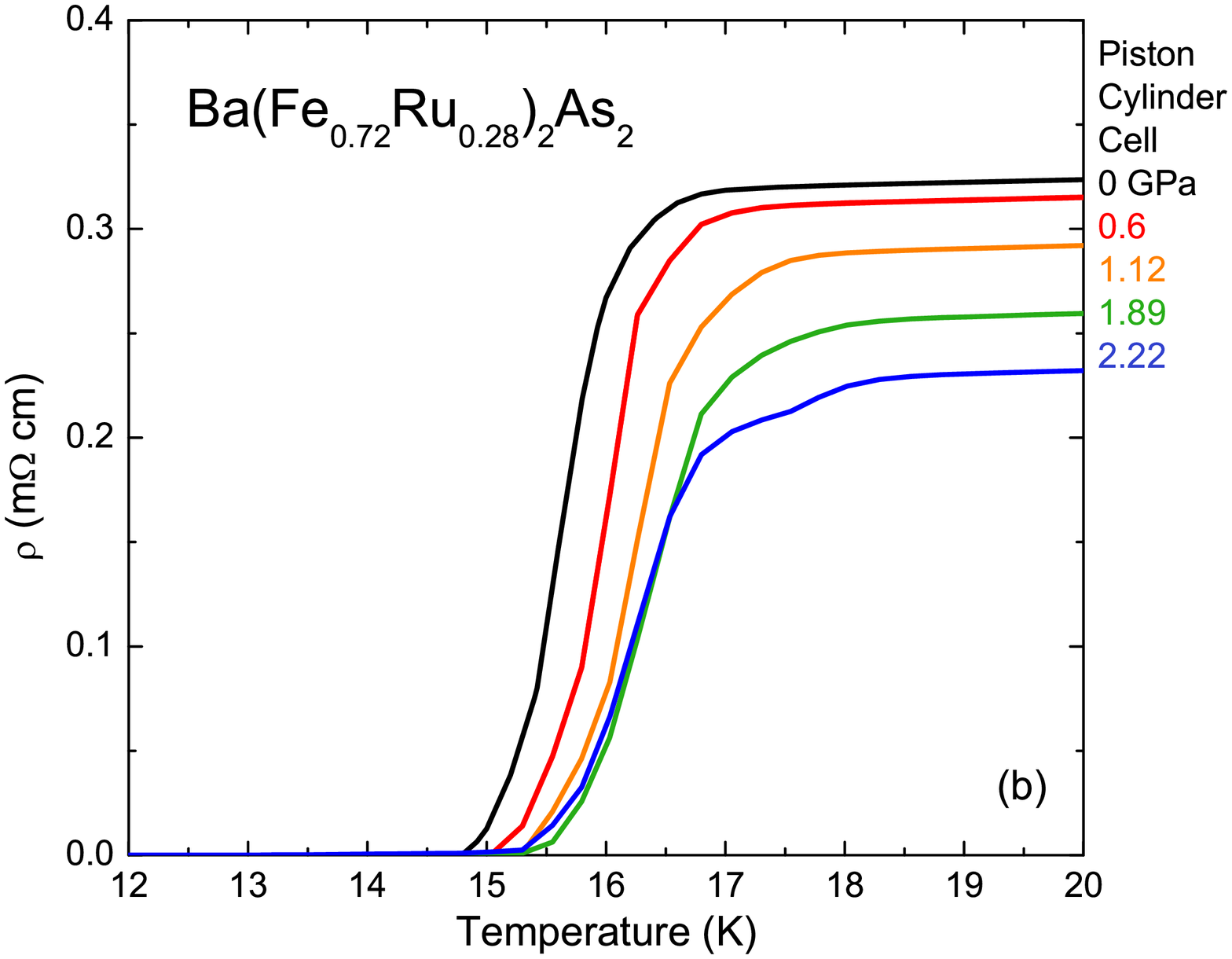}
\end{center}
\caption{(Color online) Temperature dependence of the resistivity of \BaRuTwoEight up to 2.22~GPa measured using the piston cylinder cell (a)~Shown for temperatures up to 300~K. (b)~Shown for temperatures up to 20~K.}
\label{Ru28}
\end{figure}

\begin{figure}[!ht]
\begin{center}
\includegraphics[angle=0,width=90mm]{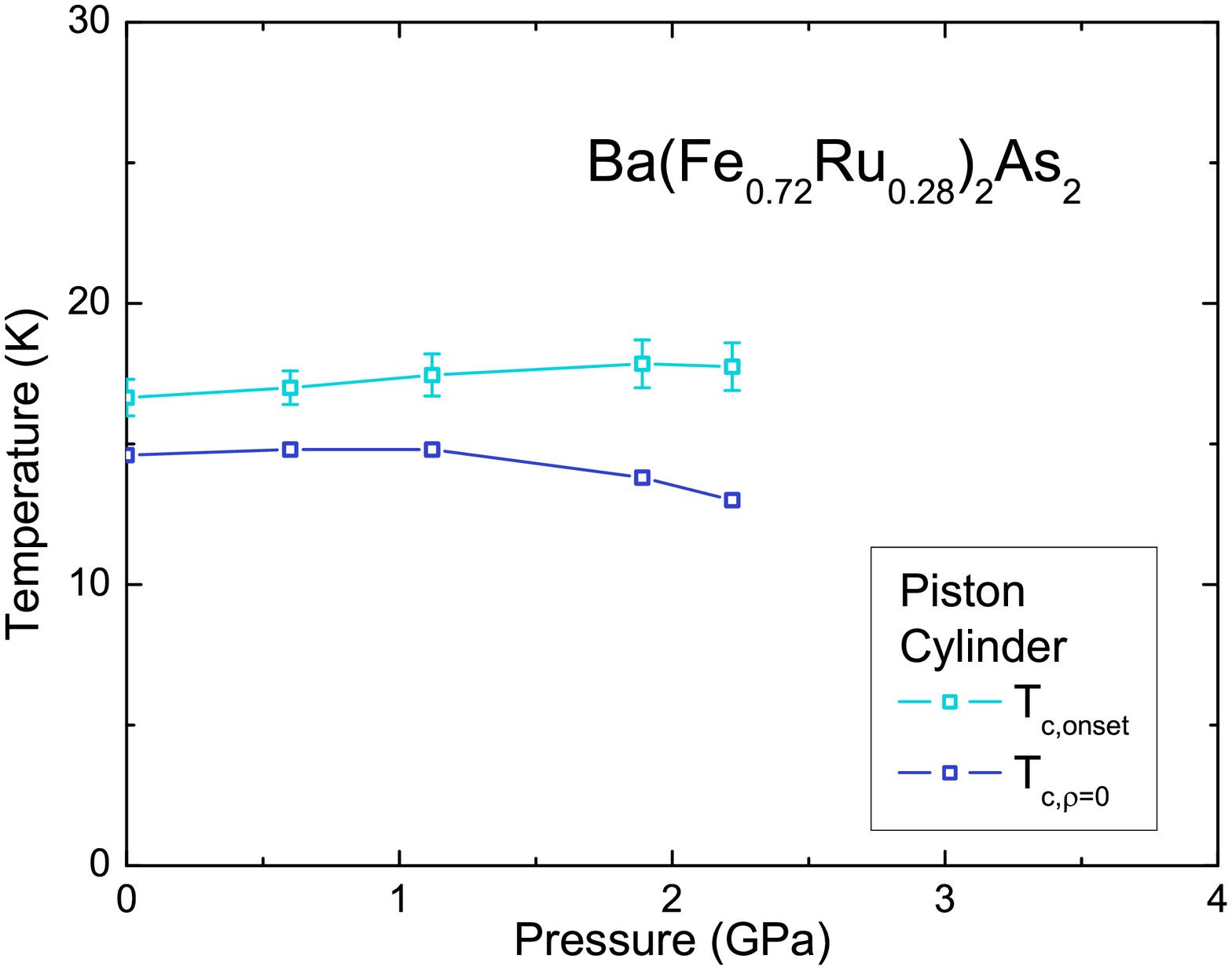}
\end{center}
\caption{(Color online) $T-P$ Phase Diagram for measurements of \BaRuTwoEight using the piston cylinder cell.}
\label{Ru28PD}
\end{figure}

\BaRuTwoEightf , having optimal Ru concentration, shows no structural/magnetic transition at ambient pressure and the superconducting transition is relatively sharp with \Tcf ~$\sim$~16~K and the transition width $\Delta T_{c}~\sim$~0.7~K.  Added pressure marginally increases $T _{c,onset}$ and in fact widens the transition width with $T _{c,\rho =0}$ decreasing as shown in Figs.~\ref{Ru28} and \ref{Ru28PD}. The superconducting onset and offset temperatures show very little scatter compared to low pressure measurements on the other Ru substituted samples.  

\section{Discussion}
Previous pressure studies have shown that \BaP and related compounds are sensitive to the degree of non-hydrostaticity of the pressurized environment.\cite{Colombier09,Matsubayashi09,Ishikawa09,Duncan10,Yamazaki10}  Empirically, increasingly hydrostatic environments move \Tsm and \Tc to higher pressures on the $T-P$ phase diagram.  Having the pressure-transmitting medium still be a liquid at room temperature when pressure is increased reduces the degree of uniaxial stress on the sample.  In such cases any non-hydrostaticity is caused by the differential thermal contractions of the various components of the cell below the vitrification/solidification temperature (the temperature below which the liquid medium changes into a glass or solid).  

For measurements taken with the piston cylinder cell, the superconducting onsets were broader and more rounded than those taken with the Bridgman cell.  This is expected since the samples for the piston cylinder cell were typically twice as long as those for the Bridgman cell.  Longer samples are more vulnerable to pressure inhomogeneities due to the larger region across which strain gradients can occur. 

\begin{figure}[!ht]
\begin{center}
\includegraphics[angle=0,width=100mm]{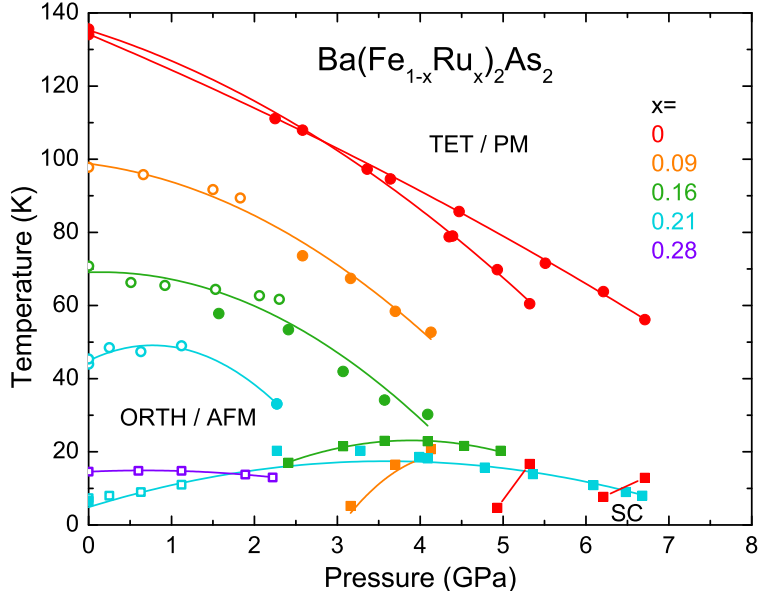}
\end{center}
\caption{(Color online) Combined phase diagram for all Ru concentrations.  Open circles and squares are \Tsm and \Tcf , respectively, from piston cylinder cell measurements.  Solid circles and squares are \Tsm and \Tcf , respectively, from Bridgman cell measurements.  Lines are guides for the eyes.}
\label{AllPD}
\end{figure}
   
The effect of strain gradients on the samples can also be invoked to explain the relatively low pressure sensitivity of the $T_{c,onset}$ line in the phase diagrams.  Given that $T_{c,\rho =0}$ forms a fairly well defined, pressure dependent dome-like region,  $T_{c,onset}$ can be understood in terms of an effective strain gradient over some region of the sample, equivalent to several GPa.  With such a gradient, a wide distribution of the \Tc values could exist leading to a fairly pressure insensitive $T_{c,onset}\sim T_{c,max}$.  This is precisely what is seen here as well as in SrFe$_{2}$As$_{2}$ \cite{Saha09} and inferred by Nakashima, \textit{et al.}\cite{Nakashima10}  Based on this premise, we pay far greater attention to the $T_{c,\rho =0}(P)$ data.  

All of the $T-P$ phase diagrams for \BaRu are shown together in Fig.~\ref{AllPD}.  Although the suppression of \Tsm with increasing Ru concentration and pressure is clear, as is the stabilizing of the superconducting region, this plot does not clearly reveal any other unifying trends.  

\begin{figure}[!ht]
\begin{center}
\includegraphics[angle=0,width=90mm]{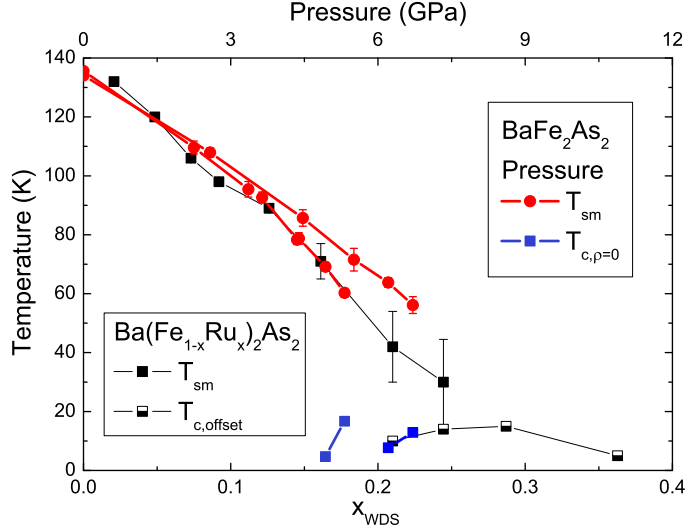}
\end{center}
\caption{(Color online) Comparison of the $T-x$ phase diagram for \BaRu and $T-P$ phase diagram of \BaP with a ratio of 3~GPa to $x$~=~0.10~Ru relating the two horizontal axes.}
\label{PDOverlay}
\end{figure}

In the earlier study of Ru substitution in \BaPf ,\cite{Thaler10} a comparison was made between the $T-x$ phase diagram and the $T-P$ phase diagram of the parent compound.  We make the same comparison here, in Fig.~\ref{PDOverlay}, with measurements taken with the iso-pentane~:~n-pentane mixture.  Although the full superconducting dome was not determined under pressure for pure \BaPf , by overlapping the \Tsm suppression curve, it is readily seen that 3~GPa is grossly comparable to $x$~=~0.10~Ru substitution for these pressure conditions.  It should be noted that for the Fluorinert 70~:~77 pressure medium used in the Bridgman cell reported by Thaler, \textit{et al.},\cite{Thaler10,Colombier09} this relation was close to 2~GPa to $x$~=~0.10~Ru.  Clearly this relationship depends on multiple factors, most likely associated with non-hydrostatic pressure components due to the freezing of the liquid medium.

\begin{figure}[!ht]
\begin{center}
\includegraphics[angle=0,width=100mm]{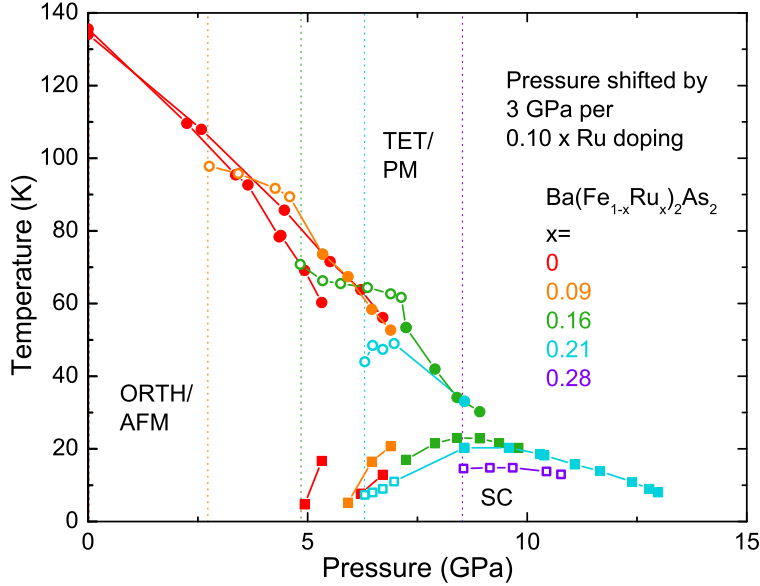}
\end{center}
\caption{(Color online) Phase diagram of all Ru concentrations each shifted by 3~GPa for every $x$~=~0.10~Ru substitution.  Open circles and squares are \Tsm and \Tcf , respectively, from piston cylinder cell measurements.  Solid circles and squares are \Tsm and \Tcf , respectively, from Bridgman cell measurements.}
\label{MovedPD}
\end{figure}

Using this relation from pure \BaP under pressure and ambient pressure \BaRuf , a more revealing, composite phase diagram can be created by shifting the $T-P$ phase diagrams for the various Ru concentrations according to the ratio 3~GPa~:~$x$~=~0.10~Ru.  When this is done (see Fig.~\ref{MovedPD}), the data form a much more consistent picture with \Tsm and \Tc manifolds lying roughly on top of each other.  It is important to point out that although the pressure~:~Ru~concentration ratio was based on \Tsm normalization, the $T_{c,\rho =0}$ data fall onto a consistent manifold as well.  Figure \ref{MovedPD} demonstrates that for all Ru concentrations that were studied, only a single scaling, 3~GPa for $x$~=~0.10~Ru, is necessary to line up the phase diagrams. This means that the effects of pressure and Ru substitution on \BaP are additive in a simple manner across the whole phase diagram.  Whereas both pressure and Ru substitution are nominally isoelectronic, a similar composite phase diagram can be assembled from $T-x-P$ data collected on Ba(Fe$_{1-x}$Co$_{x}$)$_{2}$As$_{2}$ samples.\cite{Colombier10}  In this non-isoelectronic case, a scaling of 0.8~GPa~:~$x$~=~0.01~Co gives the best collapse of the data onto single \Tsm and \Tc manifolds.  This result implies that the additive nature of doping and pressure may not be limited to isoelectronic substitutions.  

\begin{figure}[!ht]
\begin{center}
\includegraphics[angle=0,width=100mm]{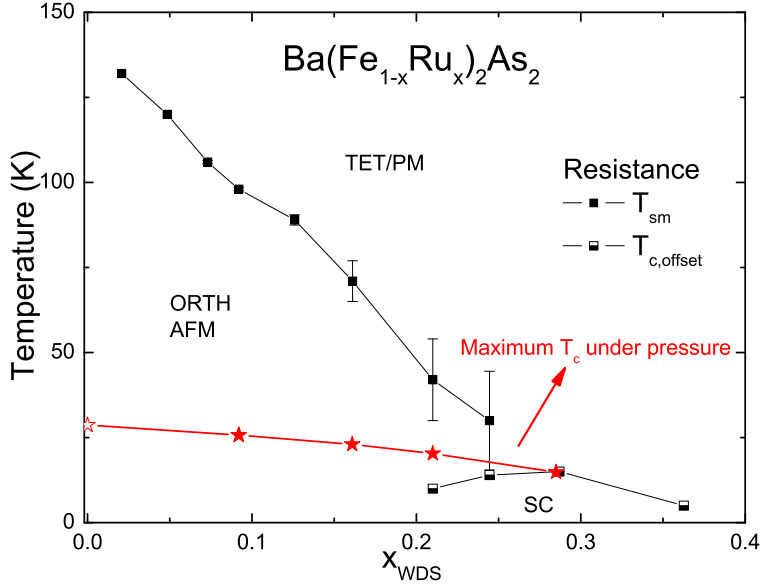}
\end{center}
\caption{(Color online) Comparison of $T-x$ phase diagram and the maximum $T_{c,\rho =0}$ achieved with pressure at various Ru concentrations. Solid stars are $T_{c,max}$ values from this study.  The open star is the $T_{c,max}$ reported by Colombier, \textit{et al.}\cite{Colombier09}}
\label{Tcmax}
\end{figure}

Another way of seeing the effect of pressure on the \BaRu system, is to plot the maximum $T_{c,\rho =0}$ on the ambient pressure $T-x$ phase diagram, (Fig.~\ref{Tcmax}). Because $P_{crit}$ was not reached with parent \BaPf , with this pressure medium, we use the maximum value reported by Colombier, \textit{et al.}\cite{Colombier09} as an estimate.  For the lower-than-optimal Ru substituted samples, as pressure suppresses the structural/magnetic transition, \Tc dramatically increases, as was the case for Co substituted \BaPf .\cite{Colombier10}  On the other hand, if \Tsm has already been suppressed, by either Co or Ru substitution, pressure no longer increases \Tcf , but rather suppresses it.  This is consistent with the idea that long range structural/magnetic ordering is detrimental for superconductivity and is the primary reason \Tc is low or zero in sub-optimally substituted samples.  

Figures \ref{MovedPD} and \ref{Tcmax} bring up an interesting question, perhaps a key one: Do Ru substitution and pressure produce similar phase diagrams via similar or different mechanisms?  At a gross level the reason for the similarity is the same: Suppression of \Tsm leads to an increase in \Tc (as has been observed for a wide range of transition metal substitutions\cite{Canfield10}).  Both Ru substitution and pressure suppress \Tsmf ; the question becomes whether this is accomplished via similar or different mechanisms.  Whereas it is fairly certain that pressure can only change details of the band structure (such as nesting) Ru substitution may change the band structure \cite{Rullier10} or it may suppress the magnetic transition temperature by replacing Fe with a far less magnetic ion.  This would be a less dramatic example of substituting Y or Lu for R~=~Gd~-~Tm in a rare earth intermetallic compound,\cite{Wiener00} perhaps involving Stoner enhancement, rather than local moments.  Ultimately, systematic studies, across the whole Ru series, via ARPES will help address these questions.

\section{Conclusion}

Pressure measurements have been carried out on the \BaRu system.  The resulting phase diagrams show a suppression of \Tsm and an enhancement of \Tc up to $P_{crit}$ where we see the narrowest superconducting transition, $T_{c,max}$, and the disappearance of \Tsm by the addition of pressure for under-doped compounds.  For the optimal Ru concentrations, further pressure increases beyond $P_{crit}$ lowers \Tc and broadens the superconducting transition.  Comparisons between the \BaRu $T-x$ phase diagrams indicate an additive correlation between physical pressure and Ru substitution of 3~GPa to $x$~=~0.10~Ru concentration.  A comparison between $T_{c,max}$ and the $T-x$ phase diagram indicate that suppression of the structural/magnetic transition is necessary for superconductivity to reach its maximum \Tc values. 

\begin{acknowledgements}
We thank E. D. Mun, X. Lin, and A. Kreyssig for enlightening discussions. This work was carried out at Ames Laboratory, US DOE, under contract \# DE-AC02-07CH11358 (SKK, EC, AT, SLB and PCC). Part of this work was performed at the Iowa State University and supported by the AFOSR-MURI grant \# FA9550-09-1-0603 (MST and PCC). MST was supported in part by the National Science Foundation under Grant \# DMR-0805335. SLB acknowledges partial support from the State of Iowa through Iowa State University.
\end{acknowledgements}

\appendix*
\section*{Appendix}
\begin{figure}[!ht]
\begin{center}
\includegraphics[angle=0,width=100mm]{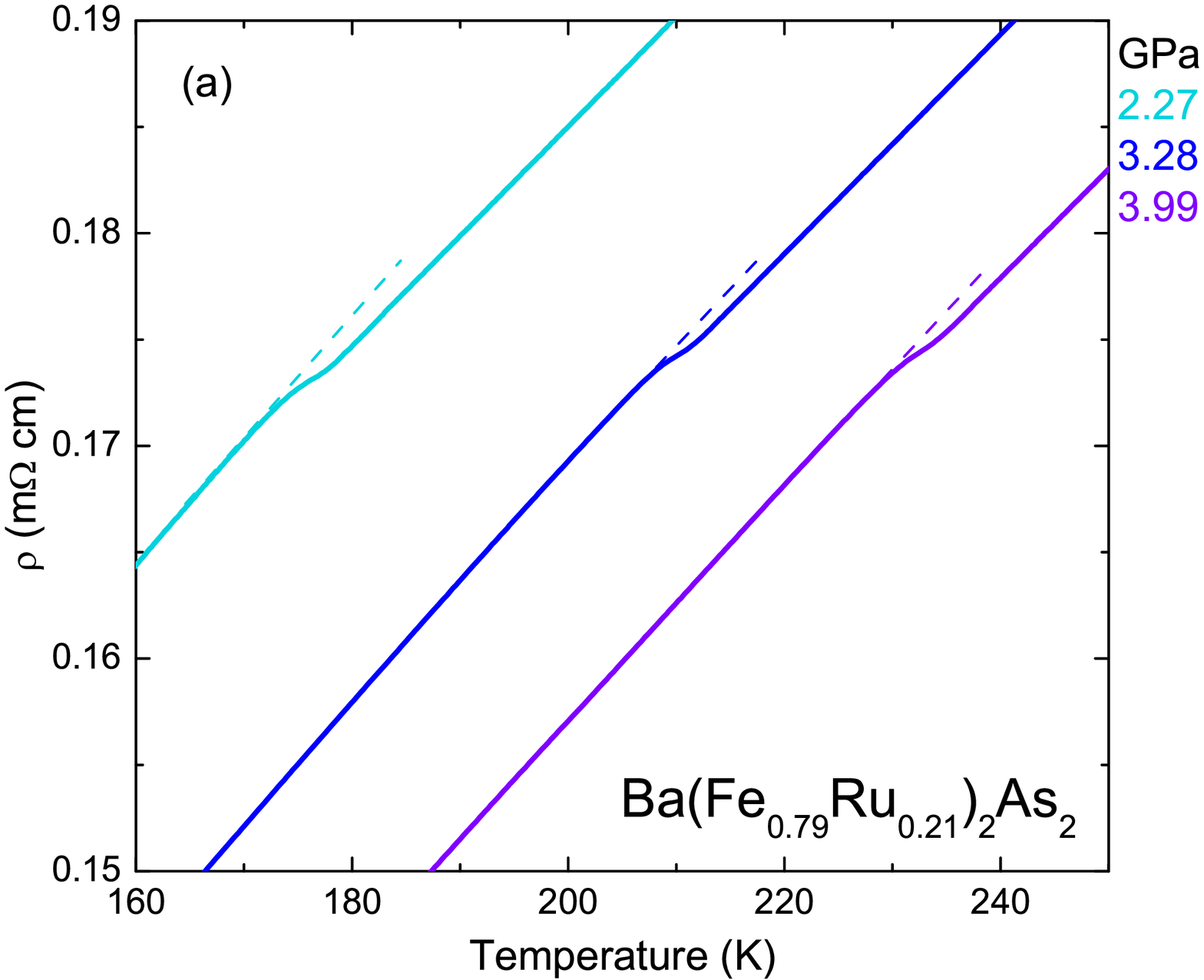}
\includegraphics[angle=0,width=100mm]{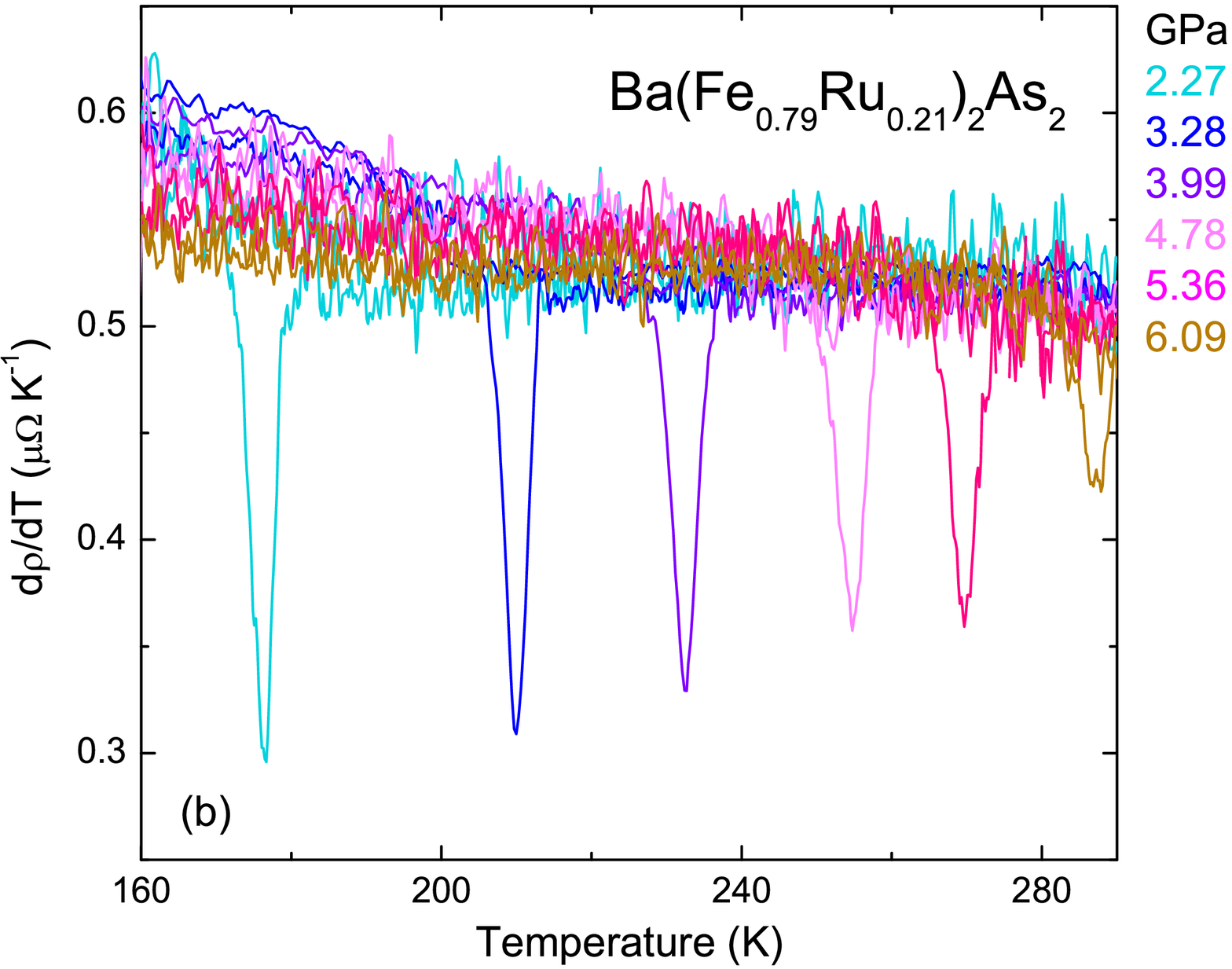}
\end{center}
\caption{(Color online) (a)~Feature in resistivity data for Ba(Fe$_{0.79}$Ru$_{0.21}$)$_{2}$As$_{2}$ at 2.27, 3.28, and 3.99~GPa. Dashed are extrapolations of the lower temperature, linear $\rho(T)$ data.  (b)~Feature in $d\rho/dT$ indicative of the melting of the liquid medium.}
\label{Melting}
\end{figure}

Given the importance of hydrostaticity for the measurements of $\rho (T)$ under pressure, we opted for a pressure medium that solidifies at relatively high pressure at ambient temperature for each pressure cell, thus reducing non-hydrostatic components associated with the pressurization process.  As a side product of this study, we were able to use the sensitivity of \BaRu to pressure conditions to track the melting temperature of the two liquid media at various pressures.  It was found that on warming, the resistivity data for various pressures and samples showed a small, anomalous, kink-like feature at higher temperatures (see Fig.~\ref{Melting}(a)).  Because this resistive anomaly consistently appeared at similar temperatures for similar pressures, and was independent of Ru content, it was attributed to a subtle change in the pressure conditions.  For the Bridgman cell with the 1~:~1 iso-pentane~:~n-pentane mixture, this feature was found to correspond to the melting temperature of the liquid medium.\cite{Klotz06}  Although this feature is essentially invisible in the $\rho(T)$ plots shown in the main text and is even difficult to see in the expanded Fig.~\ref{Melting}(a), this feature is readily seen in the derivative of the resistivity, Fig.~\ref{Melting}(b).  The minimum of this derivative was taken as the vitrification/solidification temperature of the liquid medium.

\begin{figure}[!ht]
\begin{center}
\includegraphics[angle=0,width=100mm]{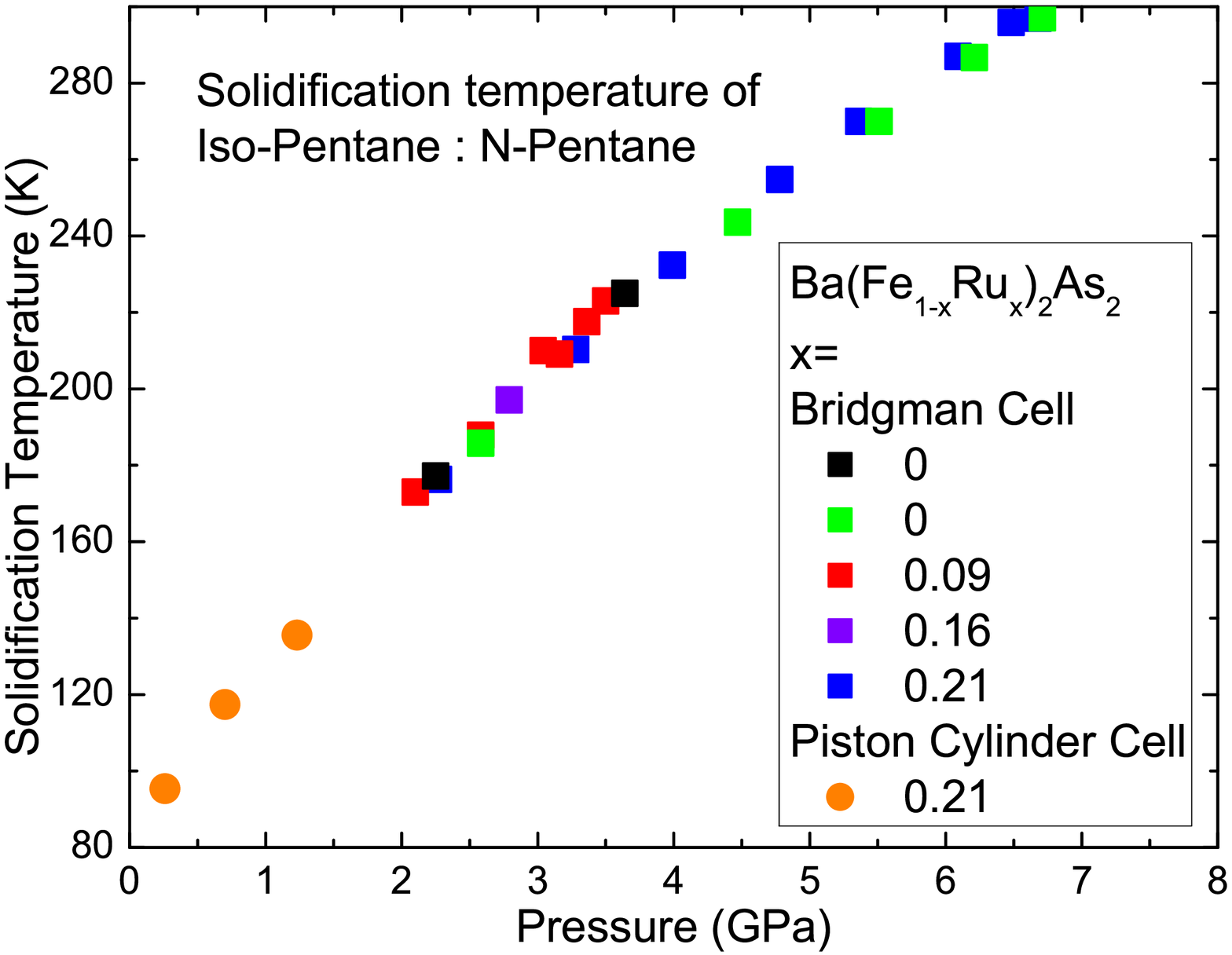}
\end{center}
\caption{(Color online) Combined phase diagram of the melting of the liquid medium (1~:~1 iso-pentane~:~n-pentane).}
\label{MeltingPD}
\end{figure}

The $T-P$ phase diagram inferred from these data is presented in Fig.~\ref{MeltingPD}.  When this curve is extrapolated to zero pressure, the melting event occurs at $\sim$~85~K.  This is lower than the previously reported freezing temperatures (105~K and 125~K at ambient pressure\cite{Klotz06,Sundqvist87}) but this discrepancy is not unexpected given the different criterion used to determine this: Sundqvist, \textit{et al.}\cite{Sundqvist87} measured the resistivity of a Manganin wire, suspended in this liquid medium and noted the temperature at which the resistivity dramatically diverges from the expected linear behavior, indicating the onset of solidification; on the other hand, Klotz, \textit{et al.}\cite{Klotz06} used the ``blocked-capillary method" where a thin capillary inside a temperature-controlled, copper block is filled with the liquid medium.  In this case, the reported values are for temperatures where the liquid medium attains a viscosity similar to thick molasses.  

More importantly, it is useful to know the hydrostatic limit of the liquid medium at the temperature when pressure is applied; usually this is at room temperature ($\sim$~300~K).  Both Piermarini, \textit{et al.}\cite{Piermarini73} and Klotz, \textit{et al.}\cite{Klotz09} placed rubies in a diamond anvil cell filled with the iso-pentane and n-pentane mixture.  At 7.4~GPa, they saw a broadening of the spectral line of rubies that could be correlated to the solidification of the medium.  The hardness of rubies makes them less sensitive to pressure gradients, therefore 7.4~GPa can be considered a higher hydrostatic limit of the liquid medium.  A different approach was used by Nomura \textit{et al.}\cite{Nomura82} where, once again, the resistivity of a Manganin wire was suspended in the liquid medium, but this time, inside a cubic anvil pressure cell.  At 283~K, the resistivity of the Manganin wire diverged from the expected linear behavior at 5.6~GPa.\cite{Nomura82}  

In our study, the the anomaly seen in the resistivity curves indicate that the melting event occurs at $\sim$~6.5~GPa at 300~K which is within the range of previously reported values. In fact at $\sim$~283~K, the hydrostatic limit from our study is 6.0~GPa which is only 0.4~GPa higher than the results from Nomura, \textit{et al.}\cite{Nomura82}  

The advantage of this study was that the freezing transition was tracked across a wide range of temperatures and pressures.  Previous reports\cite{Klotz09,Klotz06,Piermarini73,Nomura82} on the vitrification/solidification of the iso-pentane~:~n-pentane mixture were typically studied only at a given temperature or pressure.

\begin{figure}[!ht]
\begin{center}
\includegraphics[angle=0,width=100mm]{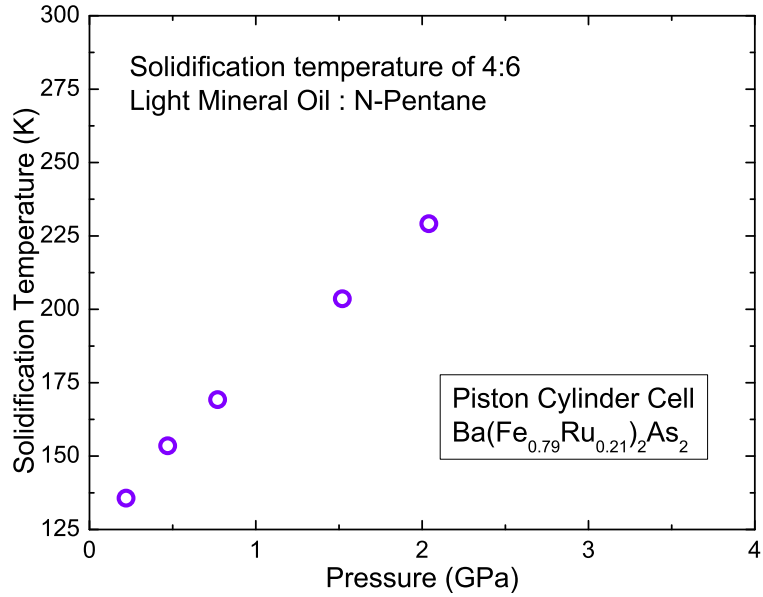}
\end{center}
\caption{(Color online) Combined phase diagram of the melting of the liquid medium 4~:~6 light mineral oil~:~n-pentane.}
\label{MeltingPD2}
\end{figure}

In a similar manner, the vitrification/solidification temperature of the 4~:~6 light mineral oil~:~n-pentane mixture was determined at several pressures using the piston cylinder cell.  The resistivity data for \BaRuTwoOne taken on warming with this liquid medium showed a similar anomalous kink the the derivative of the resistivity data showed a clear feature that we took to be the vitrification/solidification event.  Figure \ref{MeltingPD2} shows the $T-P$ phase diagram for this liquid medium.  At 300~K, this phase transition is expected to occur at a pressure of roughly 3.5~GPa, thus quantitatively justifying the use of this liquid medium at pressures up to 3~GPa in the past.\cite{Budko84}

\end{document}